\documentclass[twocolumn,pra,amsmath,amssymb,floatfix,superscriptaddress,nofootinbib,showpacs]{revtex4-2}
\usepackage{amsmath,amssymb,graphicx,epsfig}
\usepackage{bbm}
\usepackage{color}
\usepackage{mathtools}
\usepackage{mathrsfs}

\makeatletter
\let\MYcaption\@makecaption
\makeatother

\usepackage[font=footnotesize,labelformat=simple]{subcaption}

\makeatletter
\let\@makecaption\MYcaption
\makeatother

\DeclarePairedDelimiter\bra{\langle}{\rvert} 
\DeclarePairedDelimiter\ket{\lvert}{\rangle} 
\DeclarePairedDelimiterX\braket[2]{\langle}{\rangle}{#1 \delimsize\vert #2} 

\DeclareMathOperator{\Tr}{Tr}

\DeclareMathOperator{\cov}{cov}
\newcommand{\dd}{\mathrm{d}}
\newcommand{\btau}{\bar{\tau}}
\newcommand{\bs}{\bar{s}}

\begin{document}
\title{Disentanglement approach to quantum spin ground states: field theory and stochastic simulation}
\author{Stefano De Nicola}
\affiliation{IST Austria, Am Campus 1, 3400 Klosterneuburg, Austria}

\begin{abstract}
While several tools have been developed to study the ground state of many-body quantum spin systems, the limitations of existing techniques call for the exploration of new approaches. In this manuscript we develop an alternative analytical and numerical framework for many-body quantum spin ground states, based on the disentanglement formalism. In this approach, observables are exactly expressed as Gaussian-weighted functional integrals over scalar fields. We identify the leading contribution to these integrals, given by the saddle point of a suitable effective action. Analytically, we develop a field-theoretical expansion of the functional integrals, performed by means of appropriate Feynman rules. The expansion can be truncated to a desired order to obtain analytical approximations to observables. Numerically, we show that the disentanglement approach can be used to compute ground state expectation values from classical stochastic processes. While the associated fluctuations grow exponentially with imaginary time and the system size, this growth can be mitigated by means of an importance sampling scheme based on knowledge of the saddle point configuration. We illustrate the advantages and limitations of our methods by considering the quantum Ising model in 1, 2 and 3 spatial dimensions. Our analytical and numerical approaches are applicable to a broad class of systems, bridging concepts from quantum lattice models, continuum field theory, and classical stochastic processes. 
\end{abstract}

\maketitle

\section{Introduction}
Lattice quantum spin systems constitute an important class of models in many-body physics. 
Spin Hamiltonians represent different universality classes in condensed matter systems \cite{Blundell2001,Sandvik2010,Grosso2014}. 
More recently, advancements in the fields of ultracold atomic gases \cite{reviewColdAtoms2015,Gross2017} and trapped ions \cite{blattRoos2012,schneider2012ions} have made it possible to experimentally realize isolated model systems, enabling a direct investigation of fundamental concepts such as quantum phase transitions and entanglement.
Away from exactly solvable integrable models \cite{korepin_bogoliubov_izergin_1993,gaudin_2014}, which are mostly one dimensional, analytical treatments of quantum spin systems are typically based on the spin coherent state path integral \cite{Klauder1979,Berezin1980,Perelomov1986,Auerbach1994,AltlandSimons2010,Fradkin2013}. While path integrals frequently elude an exact evaluation, they are often useful to develop approximations, including semiclassical treatments and instanton techniques \cite{Auerbach1994,AltlandSimons2010,Fradkin2013}.
However, the continuum limit of the coherent state path integral is mathematically subtle~\cite{solari1987,Kochetov1998,Stone2000,Wilson2011}, and is still an area of current research~\cite{Wilson2011,kordas2014,KORDAS2016226,Taniguchi2017,Kochetov2019}. When taking the continuum limit, the differentiability of trajectories is incorrectly assumed~\cite{Klauder1979,Auerbach1994,ringelGritsev}, which can produce wrong results even for simple models~\cite{Wilson2011}.
The lack of generally applicable analytical techniques has also led to the development of several numerical schemes.
Quantum Monte Carlo methods \cite{Sandvik2010,Avella2013} have achieved great success for a range of systems \cite{Creswick1988,Beard1998,Aplesnin1998,Blote2002,JiangNyfeler2008,Lou2009,Sandvik2010,%
Shao2016,Zhao2019,Hen2019}; other applications (notably, frustrated magnets \cite{lacroix2011introduction}) are however plagued by sign or phase problems~\cite{Henelius2000,Troyer2005,marvian2019}, which have been circumvented in special cases \cite{Nakamura1998,Moessner2001,Isakov2006,
Kaul2013,Alet2016,hann2017,Wessel2017}, but whose general resolution has proved to be a hard task.
More recent tensor network approaches~\cite{Schollwock2011,Orus2014} have been able to handle large or even infinite systems in one~\cite{Vidal2006} and higher dimensions~\cite{verstraete2004,Jordan2008,Jiang2008,Lubasch2014}; however, their applicability in the latter case is significantly limited by the growth of entanglement and the computational cost associated with contracting higher dimensional lattices \cite{verstraete2006,schuch2007,Lubasch2014}. 

An alternative framework for quantum spin systems has recently emerged, based on a disentanglement approach~\cite{hoganChalker,galitski,ringelGritsev} whereby quantum expectation values are exactly expressed as functional integrals over single-spin trajectories. 
Said integrals are performed with respect to the Wiener measure~\cite{kloeden} and are thus straightforwardly amenable to numerical evaluation~\cite{ringelGritsev,stochasticApproach}. Furthermore, as noted in Ref.~\cite{ringelGritsev}, this construction does not assume the differentiability of paths, and is therefore free from the related issues that affect coherent state path integrals~\cite{Klauder1979,Auerbach1994,ringelGritsev}.
The field theoretical description provided by the disentanglement formalism has been used to obtain analytical results for certain integrable models~\cite{hoganChalker,ringelGritsev}. However, a generally applicable analytical method to compute observables from this approach is presently lacking.
Alternatively, the path integrals obtained from the disentanglement method can be evaluated by numerically solving a set of stochastic differential equations~\cite{hoganChalker,ringelGritsev}.
While this approach has been recently investigated in non-equilibrium settings~\cite{stochasticApproach,nonEquilibrium}, much less is known in the context of ground states.

In this manuscript, we explore ground state applications of the disentanglement approach, developing a set of analytical and numerical tools.
In Section~\ref{sec:disentanglement}, we briefly review the disentanglement formalism and apply it in Euclidean time to exactly cast ground state expectation values in path integral form. Going beyond previous applications of the disentanglement formalism, we then identify the trajectory yielding the largest contribution to a given observable: this corresponds to the saddle point configuration extremizing a suitably defined effective action, as we illustrate for the quantum Ising model in $D$ spatial dimensions. 
On the analytical side, in Section~\ref{sec:analytical} we then show how to systematically compute corrections beyond the saddle point approximation; we provide an example for the quantum Ising chain, introducing an appropriate set of Feynman rules and diagrams.
On the numerical side, by biasing the measure towards the saddle point configuration we obtain an exact importance sampling scheme, greatly improving the performance of the method over direct sampling. We show this in Section~\ref{sec:importanceSampling}, where we compute observables for 1, 2 and 3 dimensional systems and study the behavior of fluctuations in the corresponding stochastic quantities.
Finally, in Section~\ref{sec:conclusion} we summarize our findings, discussing the advantages and limitations of the disentanglement method, and outline future directions.

\section{Disentanglement Formalism for Quantum Spin Ground States}\label{sec:disentanglement}
\subsection{Disentanglement Transformation}\label{sec:stochasticFormalism}
Consider a quantum spin system with Hamiltonian $\hat{H}$. The ground state $\ket{\psi_{G}}$ of the system can be obtained from a generic state $|\psi_0 \rangle$ by performing imaginary time evolution: since at late imaginary times $\tau$ all excited states are exponentially suppressed compared to the ground state, one has
\begin{equation}\label{eq:gs}
\ket{\psi_{G}} \sim \lim_{\tau \rightarrow \infty} e^{-\hat H \tau}|\psi_0 \rangle ,
\end{equation}
where we set $\hbar=1$. It is then natural to introduce the Euclidean time evolution operator $\hat U(\tau)=e^{- \hat H \tau }$. Without loss of generality, let us consider initial states $\ket{\psi_0}$ that are product states; these can be conveniently parameterized in terms of a single reference state $\ket{\Downarrow} \equiv \prod_i \ket{\downarrow}_i$, where $\hat{S}^-_i \ket{\downarrow}_i=0$. The Euclidean time evolution from an arbitrary state $\ket{\psi_0}$ is then obtained by considering the modified time evolution operator 
\begin{align}\label{eq:modifiedU}
\hat{\mathcal{U}}(\tau) \equiv \hat{U}(\tau) \hat{U}_0 ,
\end{align}
where the unitary operator $\hat{U}_0$ satisfies
\begin{align}\label{eq:initialState}
\hat{U}_0 \ket{\Downarrow} = \ket{\psi_0}.
\end{align}
The ground state expectation value of an observable $\hat{\mathcal{O}}$ can thus be written as 
\begin{align}\label{eq:GSexpectationValue}
\mathcal{O}_{G}= \lim_{\tau \rightarrow \infty} \frac{\bra{\Downarrow}\hat{\mathcal{U}}^\dag(\tau) \hat{\mathcal{O}} \hat{\mathcal{U}}(\tau)\ket{\Downarrow}}{\bra{\Downarrow} \hat{\mathcal{U}}^\dag(\tau) \hat{\mathcal{U}}(\tau)\ket{\Downarrow}}.
\end{align}
The denominator of Eq.~(\ref{eq:GSexpectationValue}) provides the necessary normalization, since $\hat{\mathcal{U}}(\tau)$ inherits the non-unitarity of $\hat{U}(\tau)$. All information about ground state expectation values is then encoded in the late-time behavior of $\hat{\mathcal{U}}(\tau)$.
Let us consider spin systems with a quadratic Hamiltonian
\begin{equation}\label{eq:hamiltonian}
  \hat H= - J \sum_{ijab} \mathcal{J}_{ij}^{ab}\hat S_i^a\hat S_j^b - \sum_{ia} h_i^a\hat S_i^a ,
  \end{equation}
where $a, b \in \{ +,-,z \}$ and the spin operators $\hat S^a_j$ on site $j$ satisfy the SU($2$) commutation relations $[\hat S_j^z,\hat S_{j^\prime}^\pm]=\pm \delta_{jj^{\prime}} \hat S_j^\pm $, $[\hat S_j^+,\hat S_{j^\prime}^-]=2 \delta_{jj^{\prime}} \hat S_j^z $.
We consider a symmetric interaction matrix $\mathcal{J}^{ab}_{ij}$ with interaction strength $J$ and an applied magnetic field $h^a_i$.
For systems whose Hamiltonian is of the form~(\ref{eq:hamiltonian}), the operator $\hat{\mathcal{U}}(\tau)$ can be conveniently re-expressed using a disentanglement formalism~\cite{hoganChalker, galitski, ringelGritsev}, recently applied in the context of real time evolution~\cite{stochasticApproach,nonEquilibrium}.
By performing a Hubbard-Stratonovich (HS) decoupling \cite{stratonovich,hubbard} followed by a Lie-algebraic disentanglement transformation
\cite{weiNorman, kolokolov, hoganChalker,galitski, ringelGritsev}, $\hat{\mathcal{U}}(\tau)$ can be exactly represented as a functional integral \cite{hoganChalker,galitski,ringelGritsev}:
\begin{equation}\label{eq:evOptr}
\hat{\mathcal{U}}(\tau) =  \int \mathcal{D} \varphi
\mathrm{e}^{-S_0[\varphi]} \hat{\mathcal{U}}^s(\tau),
\end{equation}
where the \textit{noise action} $S_0$ is given by\footnote{This convention differs from that of Refs~\cite{stochasticApproach,nonEquilibrium} by a rescaling of the scalar fields $\varphi^a_i$; see Appendix~\ref{app:stochasticDecoupling}.}
\begin{equation}
S_0[\varphi] \equiv \frac{J}{4} \int^{\tau}_{0} \mathrm{d}\tau^\prime \sum_{abij} (\mathcal{J}^{-1})_{ij}^{ab}\varphi^a_i(\tau^\prime) \varphi^b_j(\tau^\prime)
\label{eq:noiseAction}
\end{equation}
and the stochastic time evolution operator $\hat{\mathcal{U}}^s$ is defined as a product of on-site operators:
\begin{align}
 \hat{\mathcal{U}}^s(\tau) \equiv \prod_j \hat{\mathcal{U}}^s_j(\tau) = \prod_j  e^{\xi_j^+(\tau)\hat
  S_j^+}e^{\xi_j^z(\tau)\hat S_j^z}e^{\xi_j^-(\tau)\hat
  S_j^-}.
\end{align}
The operators $\hat{\mathcal{U}}^s_j$ have a functional dependence on the fields $\varphi=\{\varphi^a_i\}$ via the {\em disentangling variables} $\xi \equiv \{\xi_j^\nu \}$, which satisfy \cite{ringelGritsev}
\begin{subequations}\label{eq:SDEs}
  \begin{align}
  \dot\xi_j^+ & = \Phi_j^+ +\Phi_j^z \xi_j^+ -\Phi_i^- {\xi_j^+}^2 \label{eq:xip} ,\\
  \dot \xi_j^z & = \Phi_j^z -2 \Phi_j^- \xi_j^+ \ ,\\
  \dot \xi_j^- & = \Phi_j^- \exp\xi_j^z ,
  \end{align}
\end{subequations} 
where $\Phi_j^a=h_j^a+ J\varphi^a_j$. The initial conditions of the disentangling variables are determined from $\hat{\mathcal{U}}(0)=\hat{U}_0$; for example, for a spin-1/2 system, the general product state $\ket{\psi_0}\equiv \prod_i (a_i,b_i)$ corresponds to the initial conditions 
\begin{subequations}\label{eq:initialConditions}
\begin{align}
\xi^+_i(0)&=a_i/b_i , \\
 \xi^z_i(0)& = -2 \log(b_i), \\
 \xi^-(0)&= -a_i^*/b_i^*. 
\end{align}
\end{subequations}
For completeness, we outline the derivation of Eqs~(\ref{eq:evOptr}) and (\ref{eq:SDEs}) in Appendix~\ref{app:stochasticDecoupling}. Eq.~(\ref{eq:evOptr}) can be seen as an exact path integral representation of the time-evolution operator. The operators inside the functional average~(\ref{eq:evOptr}) are decoupled over sites and act in a simple way on any state of interest. While each trajectory parameterized by $\xi^a_i$ describes a non-interacting spin and has no entanglement, the effect of interactions is encoded in the correlations between the fields $\varphi^a_i$, determined by the noise action (\ref{eq:noiseAction}); the interacting quantum spin system is thus fully retrieved upon performing the functional integral in (\ref{eq:evOptr}).
Eq.~(\ref{eq:evOptr}) allows one to formulate an exact field theoretical description of lattice spin systems, as we show in Section~\ref{sec:analytical}. The noise action~(\ref{eq:noiseAction}) can be diagonalized in terms of a new set of fields $\phi = \{\phi^a_i\}$ by performing the linear transformation $\phi^a_i= \sum_{bj} O^{ab}_{ij} \varphi_j^b$, where $O$ is a matrix satisfying $O^T \mathcal{J}^{-1} O /2J = \mathbbm{1}$. With this transformation, the noise action takes the form \cite{stochasticApproach}
\begin{equation}
S_0[\phi] \equiv \frac{1}{2} \int^{\tau}_{0} \mathrm{d}\tau^\prime  \sum_{ai} \phi^a_i(\tau^\prime) \phi^a_i(\tau^\prime).
\label{eq:noiseActionDiagonal}
\end{equation}
Notably, due to the Gaussian nature of~(\ref{eq:noiseActionDiagonal}), the fields $\phi$ can also be interpreted as delta-correlated, unit-variance Gaussian white noise variables \cite{hoganChalker,ringelGritsev}. Thus, the functional integral in Eq.~(\ref{eq:evOptr}) can be equivalently seen as an average over the stochastic processes $\phi_i^a(\tau)$ \cite{hoganChalker,ringelGritsev,stochasticApproach}:
\begin{equation}\label{eq:stochasticTimeEvol}
\hat{\mathcal{U}}(\tau)=\big\langle  \prod_j  e^{\xi_j^+(\tau)\hat
  S_j^+}e^{\xi_j^z(\tau)\hat S_j^z}e^{\xi_j^-(\tau)\hat
  S_j^-} \big\rangle_\phi .
\end{equation}
The notation $\langle\dots \rangle_\phi$ denotes averaging with respect to the noise action~(\ref{eq:noiseActionDiagonal}), so that the relevant probability law is the Wiener measure. The equations of motion Eq.~(\ref{eq:SDEs}) are then interpreted as stochastic differential equations (SDEs) for the variables $\xi^a_i$ \cite{hoganChalker,ringelGritsev}. 
By representing each of the time evolution operators in Eq.~(\ref{eq:GSexpectationValue}) using the disentangling formula~(\ref{eq:stochasticTimeEvol}), one can express quantum ground state expectation values as classical averages. 
Introducing independent sets of forwards and backwards fields (labeled by the additional subscripts $f,b$), $\phi_f \equiv \{ \phi^a_{f,i} \}$ and $\phi_{b}\equiv \{ \phi^a_{b,i} \}$, and the associated disentangling variables $\xi_f \equiv \{ \xi^a_{f,i} \}$, $\xi_b \equiv \{ \xi^a_{b,i} \}$, one has
\begin{align}\label{eq:GSexpectationValueStochastic}
\mathcal{O}_{G} = \lim_{\tau \rightarrow \infty} \frac{\langle F_\mathcal{O}(\tau) \rangle_{\phi_f,\phi_b}}{\langle F_{\mathbbm{1}}(\tau) \rangle_{\phi_f,\phi_b}} ,
\end{align}
where the classical function $F_\mathcal{O}$ corresponding to the operator $\hat{\mathcal{O}}$ is defined by
\begin{align}\label{eq:classicalFunction}
F_\mathcal{O} \equiv \bra{\Downarrow}[\hat{\mathcal{U}}^s(\tau)]^\dag \hat{\mathcal{O}} \hat{\mathcal{U}}^s(\tau)\ket{\Downarrow},
\end{align}
and $F_\mathcal{\mathbbm{1}}$ is obtained from (\ref{eq:classicalFunction}) when $\hat{\mathcal{O}}$ is replaced by the identity operator. It can be readily seen that the functions $F_{\mathcal{O}}$ take the same functional form as their real time counterparts, given in Refs~\cite{stochasticApproach,nonEquilibrium}. Here, in contrast to said references, we express all initial states in terms of a single reference state and variable initial conditions $\xi^a_i(0)$: in this way, there is a one-to-one correspondence between observables and their classical counterparts, regardless of the initial state. The reference state $\ket{\Downarrow}$ was selected because it results in the simplest classical expressions \cite{stochasticApproach,nonEquilibrium}.
For example, for spin-$1/2$ systems one has $\hat{S}^a_i = \sigma^a_i/2$, where $\sigma^a_i$ are the Pauli matrices. In this case, by acting with~(\ref{eq:stochasticTimeEvol}) on the reference state $\ket{\Downarrow}$, the classical function for the normalization is found to be
\begin{align}\label{eq:normalization}
F_\mathcal{\mathbbm{1}}(\tau) = \prod_i [1+\xi^+_{f,i}(\tau) \xi^{+*}_{b,i}(\tau)] e^{-\frac{1}{2}[ \xi^z_{f,i}(\tau) + \xi^{z*}_{b,i}(\tau)]} ,
\end{align}
while the longitudinal magnetization $\mathcal{M}_z = \sum_j \langle\hat{S}^z_j \rangle / N $ corresponds to the classical function \cite{stochasticApproach}
\begin{align}\label{eq:longitudinalMagnetization}
F_{\mathcal{M}_z}(\tau) =  \frac{F_\mathcal{\mathbbm{1}}(\tau)}{N} \sum_j \frac{1 - \xi^+_{f,j}(\tau) \xi^{+*}_{b,j}(\tau)}{1+\xi^+_{f,j}(\tau) \xi^{+*}_{b,j}(\tau)} .
\end{align}
For any observable, the appropriate classical function can be constructed using the building blocks provided in Ref.~\cite{nonEquilibrium}. The expectation values of functions such as~(\ref{eq:normalization}) and (\ref{eq:longitudinalMagnetization}) can be evaluated numerically by averaging them over realizations of the stochastic processes $\phi^a_i$, as done in Refs~\cite{stochasticApproach,nonEquilibrium} for real time evolution; the processes $\phi^a_i$ determine the time evolution of the variables $\xi^a_i$ via the SDEs~(\ref{eq:SDEs}).
The key towards both analytical and numerical developments is identifying the trajectories $\phi$ which provide the largest contribution to each observable; we discuss this in the following Section.
\subsection{Extremal Trajectories}\label{sec:effectiveAction}

In the disentanglement formalism, ground state expectation values can be numerically computed by averaging classical functions over realizations of trajectories $\phi^a_i$ distributed according to the action~(\ref{eq:noiseActionDiagonal}).
In this approach, which we refer to as \textit{direct sampling}, one preferentially generates trajectories that are close to the non-interacting limit $\phi^a_i(\tau)=0$, as shown in Appendix~\ref{app:isingSDEs}. This was done for real-time evolution in Refs~\cite{stochasticApproach,nonEquilibrium}.
However, the trajectories that contribute most significantly to a given functional integral may be substantially different from the non-interacting trajectories, as also pointed out in~\cite{ringelThesis}.
Therefore, going beyond previous applications of the disentanglement formalism \cite{hoganChalker,galitski,ringelGritsev,stochasticApproach,nonEquilibrium}, we presently identify the trajectories yielding the dominant contributions. To this end, it is convenient to work with the action~(\ref{eq:noiseAction}) written in terms of the fields $\varphi^a_i$.
An observable $\mathcal{O}$ can be written as
\begin{align}\label{eq:generalObservable}
\mathcal{O}\equiv \langle \hat{\mathcal{O}} \rangle = \langle F_\mathcal{O}[\varphi] \rangle_{\varphi}
\end{align}
where $\varphi=\{ \varphi_\alpha \}$ denotes all of the HS fields: the collective index $\alpha$ runs over sites, Lie algebra generators, and sets of fields (e.g. forwards and backwards).
Eq.~(\ref{eq:generalObservable}) can be equivalently written as
\begin{equation}\label{eq:effectiveAction}
\mathcal{O} \equiv \int \mathcal{D} \varphi e^{-S_\mathcal{O}[\varphi]} ,
\end{equation} 
which defines the \textit{effective action} $S_\mathcal{O}[\varphi] \equiv  S_0[\varphi] - \log f_\mathcal{O}[\varphi]$ for the observable $\mathcal{O}$. The leading contribution to the integral~(\ref{eq:effectiveAction}) is given by the configuration $\varphi_{\text{SP}}(\tau)$ which extremizes the effective action:
\begin{align}\label{eq:saddlePointEquationGeneral}
\varphi_{\text{SP}} : \frac{\delta S_\mathcal{O}}{\delta \varphi}\Big\vert_{\varphi_{\text{SP}}} =0.
\end{align}
We refer to $\varphi_{\text{SP}}$ as the \textit{saddle point} (SP) field. 
The trajectory corresponding to the field $\varphi_{\text{SP}}$ provides the leading order (LO) approximation to a given observable, as we further discuss in Section~\ref{sec:leadingOrder}. 

One can regard  Eq.~(\ref{eq:effectiveAction}) as an exact field theoretical formulation of the lattice quantum spin system; Eq.~(\ref{eq:effectiveAction}) can then be expanded about the saddle point to analytically obtain corrections to observables beyond LO, computed by using the associated Feynman rules and diagrams. In practice, the expansion can be carried out up to a desired order, providing approximate analytical results for ground state expectation values. This approach is discussed and explicitly applied in Section~\ref{sec:analytical}.

Alternatively, one can view the disentanglement approach as a numerical tool. $\varphi_{\text{SP}}$ can then be used to perform an exact change of variables in the functional integral (\ref{eq:effectiveAction}): this amounts to preferentially sampling trajectories near $\varphi_{\text{SP}}$, which yield the largest contributions to the integral. Notably, in this approach one does not truncate to a given order in the fluctuations: the resulting equations are still exact and fully capture the corresponding quantum problem. The measure transformation amounts to an importance sampling scheme to improve the numerical efficiency of the stochastic approach, which is formally exact and whose practical accuracy is determined by the finite number of simulations one performs. The importance sampling method is discussed in Section~\ref{sec:importanceSampling}, where we show how this can be used to numerically access much larger system sizes than it is possible by sampling according to the naive measure~(\ref{eq:noiseActionDiagonal}).

In the next Section, we show how the saddle point trajectory is computed; this provides the starting point for both analytical and numerical developments.

\subsection{Ising Saddle Point Equation}\label{sec:findingSP}
We wish to identify the leading contribution to a given functional integral by solving Eq.~(\ref{eq:saddlePointEquationGeneral}). 
For definiteness, we illustrate this by considering the quantum Ising model in $D$ spatial dimensions, but the same formalism can be applied to any spin Hamiltonian of the form (\ref{eq:hamiltonian}). The quantum Ising model is given by the Hamiltonian
\begin{equation}\label{eq:Ising}
\hat H_{\rm I}= -J\sum_{\langle ij \rangle}^N \hat S_i^z\hat S_{j}^z - \Gamma \sum_{j=1}^N \hat S_j^x ,
\end{equation}
where $\langle ij \rangle$ denotes nearest-neighbor interactions. We consider a system of $N=N_1 \times \dots \times N_D$ spin-$1/2$ degrees of freedom on a $D$-dimensional hypercubic lattice, with periodic boundary conditions and ferromagnetic interactions $J>0$. We begin by considering the integrable one-dimensional case, and then generalize our results to $D>1$, where the model cannot be solved exactly. For $D=1$, the model~(\ref{eq:Ising}) reduces to the quantum Ising chain and is solvable in terms of free fermions \cite{Pfeuty1970}; this allows the exact computation of physical observables in the thermodynamic limit. In the present units, the Ising chain has a quantum phase transition (QPT) at $\Gamma=J/2$. For this model, the general result (\ref{eq:SDEs}) specializes to the Euclidean Ising SDEs~\cite{ringelGritsev,stochasticApproach,nonEquilibrium}
\begin{subequations}\label{eq:ITising}
\begin{align}
\dot{\xi}^+_i(\tau)&= \frac{\Gamma}{2} (1-{\xi^+_i}^2) +J  \xi^+_i \varphi_i  \label{eq:ITising1} , \\
\dot{\xi}^z_i(\tau) &= -\Gamma \xi^+_i  + J \varphi_i , \label{eq:ITising2}  \\
\dot{\xi}^-_i(\tau) &= \frac{\Gamma}{2}\exp{\xi^z_i}. \label{eq:ITising3} 
\end{align}
\end{subequations}
A natural choice of observable is the ground state energy density $\epsilon_G$. This can be computed using Eq.~(\ref{eq:GSexpectationValueStochastic}), according to the general formalism outlined in Section~\ref{sec:stochasticFormalism}. Alternatively, $\epsilon_G$ can also be obtained as
\begin{equation}\label{eq:gsEnergyLoschmidt}
\epsilon_G=- \lim_{\tau_f \rightarrow \infty} \frac{1}{N \tau_f} \log  A(\tau_f)  ,
\end{equation}
where we defined the \textit{Euclidean Loschmidt amplitude} $A(\tau_f)$ for the initial state $\ket{\psi_0}$ as
\begin{align}
A(\tau_f) = \bra{\psi_0} \hat{U}(\tau_f) \ket{\psi_0}.
\end{align}
By computing $\epsilon_G$ by means of Eq.~(\ref{eq:gsEnergyLoschmidt}), one only needs to consider a single time evolution operator, which corresponds to a single set of HS fields $\phi^a_i$. Thus, using Eq.~(\ref{eq:gsEnergyLoschmidt}) allows us to simplify the subsequent analytical developments for the purpose of illustrating the method. 
The same results can however be obtained from the general formalism of Section~\ref{sec:stochasticFormalism},  involving two time evolution operators; see Appendix~\ref{app:saddlePoint}. To further simplify our calculations, we choose the all-down initial state $\ket{\psi_0}=\otimes_j\ket{\downarrow}_j\equiv \ket{\Downarrow}$; in this case, the Loschmidt amplitude is given by the functional integral \cite{stochasticApproach}
\begin{align}\label{eq:loschmidtAmplitude}
A(\tau_f) = \int {\mathcal  D}\varphi\ e^{-S_0[\varphi]} e^{-\frac{1}{2}\sum_i \xi^z_i(\tau_f)} ,
\end{align}
where the equation of motion of $\xi^z_j$ is given by~(\ref{eq:ITising2}) and the initial conditions are $\xi^a_i(0)=0$. 
Following the discussion of Section~\ref{sec:effectiveAction}, we write the Loschmidt amplitude~(\ref{eq:loschmidtAmplitude}) as 
\begin{align}
A(\tau_f) = \int {\mathcal  D}\varphi\ e^{-S[\varphi]}
\end{align}
which defines the \textit{Euclidean Loschmidt action}:
\begin{widetext}
\begin{equation}\label{eq:loschmidtAction}
S[\varphi]=\frac{J}{2} \int_0^{\tau_f} \mathrm{d}\tau \left[ \frac{1}{2} \sum_{ij} (\mathcal{J}^{-1})_{ij} \varphi_i(\tau) \varphi_j(\tau)  - \frac{\Gamma}{J}  \sum_i \xi^+_i(\tau) + \sum_i \varphi_i(\tau) \right].
\end{equation}
\end{widetext}
The variables $\xi^+_i$ featured in the action~(\ref{eq:loschmidtAction}) are themselves functionals of $\varphi_i$, as determined by~(\ref{eq:ITising}). It follows that $S[\varphi]$ cannot be written in terms of a Lagrangian involving only the fields $\varphi$ and their time-derivatives, and it is thus not possible to obtain Euler-Lagrange equations in the standard way.
Rather, in order to obtain the saddle point field configuration, one directly extremizes the action (\ref{eq:loschmidtAction}) with respect to varying the field $\varphi_i$. This yields the \textit{Loschmidt saddle point equation}
\begin{equation}\label{eq:saddlePointLoschIsing}
\varphi_i(\tau^\prime)\vert_{\text{SP}} = \frac{\Gamma}{J } \sum_j \mathcal{J}_{ij} \int^{\tau_f}_0 \mathrm{d}\tau \frac{\delta \xi^+_j(\tau)}{\delta \varphi_j(\tau^\prime)}\Big\vert_{\text{SP}} - 1  ,
\end{equation}
where we used $\sum_j\mathcal{J}_{ij}=1$ $\forall$ $i$ and $\delta \xi^+_i /\delta\varphi_j \propto \delta_{ij}$. The subscript $\text{SP}$ denotes quantities that are evaluated at the saddle point. 
By varying Eq.~(\ref{eq:ITising1}) with respect to $\varphi_i(\tau^\prime)$, one obtains
\begin{equation}\label{eq:XiGBM}
 \frac{\delta \xi^+_i(\tau)}{\delta \varphi_i (\tau^\prime)}   = J \xi_i(\tau^\prime) \theta(\tau-\tau^\prime) e^{- \int_{\tau^\prime}^{\tau}\gamma_i(s)\mathrm{d}s } \equiv \Xi_{i}(\tau,\tau^\prime) ,
\end{equation}
where $\theta(\tau)$ is the Heaviside step function and we defined  
\begin{align}\gamma_i(s) \equiv \Gamma \xi_i^+(s) - J \varphi_i(s) .
\end{align}
Due to the translational invariance of the model~(\ref{eq:Ising}) and of the chosen initial state, at the saddle point all $\xi^+_i$ take the same value, $\xi^+_i \vert_{\text{SP} }\equiv \xi^+_{\text{SP}}$. From the translational symmetry of Eq.~(\ref{eq:saddlePointLoschIsing}), it also follows that $\varphi_i\vert_{\text{SP}}=\varphi_{\text{SP}}$ and $\Xi_i\vert_{\text{SP}} = \Xi_{\text{SP}}$. Hence, in the translationally invariant case the SP equation for the field $\varphi_{\text{SP}}$ simplifies to
\begin{align}\label{eq:loschTISP}
\varphi_{\text{SP}}(\tau^\prime) = \frac{\Gamma}{J} \int_0^{\tau_f}\Xi_{\text{SP}}(\tau,\tau^\prime) \dd \tau - 1 .
\end{align}
From Eqs.~(\ref{eq:loschTISP}) and~(\ref{eq:XiGBM}) one immediately obtains the boundary condition $\varphi_{\text{SP}}(\tau_f)=-1$; Eqs~(\ref{eq:loschTISP}),~(\ref{eq:XiGBM}) and~(\ref{eq:ITising}) further imply that $\varphi_{\text{SP}}(\tau^\prime) $ must remain real-valued at all times.
To the best of our knowledge, the functional equation~(\ref{eq:loschTISP}) cannot be solved analytically. However, for the computation of ground states one is only interested in the limit $\tau_f\rightarrow\infty$. A recursive numerical solution of Eq.~(\ref{eq:loschTISP}) shows that in this limit, away from a transient near $\tau=0$ and a boundary region at $\tau \lesssim \tau_f$, the SP equation is dominated by a time-independent plateau value $\phi_P$; see Appendix~\ref{app:saddlePoint}. We may then assume that late-time plateau values, denoted by a subscript $P$, dominate the integrals, and approximate 
\begin{align}\label{eq:plateauApproximation}
\int^{\tau_f}_0 \Xi_P(\tau,\tau^\prime) \approx \int^{\tau_f}_0  \theta(\tau-\tau^\prime) J \xi^+_P e^{-\gamma_{P}(\tau-\tau^\prime)} = \frac{J \xi^+_P}{\gamma_P } , 
\end{align}
where $\gamma_P \equiv \Gamma \xi^+_P  - J \varphi_P$. Convergence of the integral in Eq.~(\ref{eq:plateauApproximation}) requires $\gamma_P<0$. Assuming that this condition is satisfied, which can be self-consistently verified a posteriori, in the $\tau_f\rightarrow \infty$ limit Eq.~(\ref{eq:loschTISP}) is reduced to an algebraic equation for $\varphi_{P}$: 
\begin{align}\label{eq:plateauField}
\varphi_{P} = \frac{\Gamma \xi_P}{\Gamma \xi^+_P-J \varphi_P} - 1 .
\end{align}
In the absence of translational invariance, an analogous set of equations for the plateau fields can be obtained from (\ref{eq:saddlePointLoschIsing}).
Eq.~(\ref{eq:plateauField}) can be solved together with the condition that $\xi^+_P$ is a fixed point of the Euclidean dynamics when $\varphi=\varphi_P$, yielding four solutions:
\begin{align}\label{eq:Psolutions}
\begin{split}
& \xi^+_{P}= 
\begin{cases} 
[J-\sqrt{J^2-\Gamma^2}]/\Gamma \\
[J+\sqrt{J^2-\Gamma^2}]/\Gamma  \\
1 \\
-1 
 \end{cases} \hspace{-3mm}
\varphi_{P} =
\begin{cases}
-\sqrt{J^2-\Gamma^2}/J \\
\sqrt{J^2-\Gamma^2}/J \\
0 \\
0 \\ 
\end{cases} \\
&\gamma_{P} = 
\begin{cases} 
J \\ 
J \\
\Gamma \\ 
-\Gamma .
\end{cases} 
\end{split}
\end{align}
In order for the SP field to be real valued, the first and second solutions are only acceptable when $\Gamma<J$; they both give $\gamma_P = J$, and the corresponding $\xi^+_{P}$ are reciprocal to each other. The fourth solution is not acceptable as it gives $\gamma_P<0$: it corresponds to a maximum of the action~(\ref{eq:loschmidtAction}). We refer to the first and second saddle points as the \textit{small-$\Gamma$} SPs and to the third one as the \textit{large-$\Gamma$} saddle point, as they give the leading order contribution to the ground state energy in these limits; see Section~\ref{sec:leadingOrder} below.
Notably, the plateau values $\varphi_P$ in Eq.~(\ref{eq:Psolutions}) coincide with the effective fields acting on each spin within the mean field (MF) approximation; similarly, the disentangling variables $\xi^+_P$ parameterize the mean field ground states. For comparison, we provide details of the MF solution in Appendix~\ref{app:meanField}. 
This finding has a transparent physical interpretation: the path integral (\ref{eq:loschmidtAmplitude}) is a sum over configurations of non-interacting spins, i.e. product states, and the SP trajectory is the single such configuration which gives the best approximation to the ground state energy. The product state which best approximates a ground state is precisely given by mean field; the first and second saddle points in (\ref{eq:Psolutions}), which have opposite $\varphi_P$, arise from spontaneous symmetry breaking at the mean field level. 
This interpretation suggests that the correspondence between the saddle point configuration and mean field is general.
In order to compute ground state expectation values using the present method, it is therefore convenient to initialize the system in the MF ground state and subsequently perform imaginary time evolution towards the true ground state. 
This is tantamount to initializing the disentangling variables at their plateau values, $\xi^a_i(0)=\xi^a_P$,  which removes the initial transient behavior.
In principle, the above discussion should be repeated for every observable, since each corresponds to a different effective action and therefore to a different SP equation. However, it can be shown that the SP configuration of the Loschmidt action also extremizes the effective action for all physical observables, as obtained from the general formalism of Section~\ref{sec:stochasticFormalism}; see Appendix~\ref{app:saddlePoint}. The findings of this Section also readily generalize to higher dimensions. The SP solution corresponds to MF also for $D>1$; a detailed derivation is provided in Appendix~\ref{app:saddlePoint}. For instance, the plateau SP values for an isotropic quantum Ising model in $D$ spatial dimensions are given by
\begin{equation}\label{eq:PsolutionsD}
\begin{aligned}
&\xi^+_{P}= 
\begin{cases} 
[D J-\sqrt{D^2J^2-\Gamma^2}]/\Gamma  \\
[DJ+\sqrt{D^2J^2-\Gamma^2}]/\Gamma  \\
1 \\
-1 
 \end{cases} \hspace{-4mm} \\
& \varphi_{P} =
\begin{cases}
-\sqrt{D^2J^2-\Gamma^2}/DJ \\
\sqrt{D^2J^2-\Gamma^2}/DJ \\
0 \\
0 \\ 
\end{cases}  
\gamma_{P} = 
\begin{cases} 
DJ \\ 
DJ \\
\Gamma \\ 
-\Gamma .
\end{cases} 
\end{aligned}
\end{equation}
\section{Field Theory}\label{sec:analytical}
The disentanglement formalism provides an exact field theoretical representation of lattice quantum spin systems, which is obtained directly from the physical spin degrees of freedom and does not involve a continuum limit in space or the mapping of the quantum system to a higher dimensional classical one. The disentanglement approach also circumvents the use of coherent states, thus avoiding the related issues discussed in the Introduction. 
In this field theoretical framework, the saddle point field configurations give the leading order approximation to observables, with successive corrections corresponding to higher order terms in the expansion of the path integral~(\ref{eq:effectiveAction}) about the saddle points. These corrections account for entanglement in the system and can be computed by using a set of Feynman rules and the associated diagrammatic representation, introduced in this Section. 
This provides a method of broad applicability to obtain systematically improvable analytical approximations from the disentanglement formalism. 
We illustrate this for the ground state energy of the quantum Ising chain, whose exact solution provides a convenient benchmark; however, the proposed method does not rely on integrability or assume a specific dimensionality and is thus applicable to a wide range of models.
\subsection{Leading Order}\label{sec:leadingOrder}
We begin by considering the leading order term; this is given by the plateau field configurations~(\ref{eq:Psolutions}) obtained in Section~\ref{sec:findingSP}. For the remainder of this Section, it is convenient to initialize the disentangling variables at their plateau values; since $\xi^+_i(0)=\xi^+_P$ corresponds to the mean field ground state $\ket{\text{MF}}$, this is equivalent to expressing the ground state energy density in the thermodynamic limit as 
\begin{align}\label{eq:gsEnergyFromMF}
\epsilon_G =\lim_{\tau_f \rightarrow \infty} \lim_{N \rightarrow \infty} -\frac{1}{\tau_f N}\log \bra{\Downarrow} \hat{U}(\tau_f) \ket{\text{MF}} .
\end{align}
Normalization of the initial state also implies $\xi^z_i(0)=\log(1+\xi_P^{+2})$. 
Eq.~(\ref{eq:gsEnergyFromMF}) can thus be written as
\begin{align}\label{eq:gsEnergyFromMF_modifiedLosch}
\epsilon_G =  \lim_{\tau_f \rightarrow \infty} \lim_{N \rightarrow \infty} -\frac{1}{\tau_f N}\log \mathcal{A}(\tau_f)
\end{align}
in terms of a modified Loschmidt amplitude, given by
\begin{align}\label{eq:modifiedLoschmidtAmplitude}
\mathcal{A}(\tau_f) = \bra{\Downarrow} \hat{\mathcal{U}}(\tau_f) \ket{\Downarrow}. 
\end{align}
The modified time evolution operator $\hat{\mathcal{U}}(\tau_f)$ in Eq.~(\ref{eq:modifiedLoschmidtAmplitude}) is given by Eq.~(\ref{eq:modifiedU}) with the condition $\hat{\mathcal{U}}(0)\ket{\Downarrow} = \ket{\text{MF}}$; due to this definition, the modified Loschmidt amplitude $\mathcal{A}(\tau_f)$ corresponds to the same effective action~(\ref{eq:loschmidtAction}) as $A(\tau_f)$.
The analysis of the action~(\ref{eq:loschmidtAction}) in Section~\ref{sec:findingSP} concerns the infinite time limit and is independent of the initial conditions. In this limit, we can again assume that all integrals are dominated by the plateau values; therefore, the earlier discussion equally applies to the present case, and the two amplitudes $\mathcal{A}(\tau_f)$, $A(\tau_f)$ are dominated by the same large-$\tau_f$ plateaus. 

The leading order approximation to the ground state energy density in the thermodynamic limit can then be obtained as
\begin{align}\label{eq:leadingOrder}
\epsilon_G \approx - \lim_{\tau_f\rightarrow \infty} \lim_{N\rightarrow \infty} \frac{1}{\tau_f N} \log  \sum_{ \text{s.p.} } e^{-S_P} ,
\end{align}
where the sum runs over the different saddle points and the plateau action is given by
\begin{align}\label{eq:plateauAction}
S_P = \frac{NJ }{4}  \varphi_P^2\tau_f -\frac{\Gamma}{2} \xi_P \tau_f + \frac{J}{2} \varphi_P \tau_f =
\begin{cases}  
-\frac{N(J^2+\Gamma^2)}{4J} \tau_f \\
-\frac{N \Gamma}{2} \tau_f .
\end{cases}
\end{align}
The top solution in Eq.~(\ref{eq:plateauAction}) corresponds to the two small-$\Gamma$ SPs, while the bottom solution corresponds to the large-$\Gamma$ SP. The double degeneracy of the former SP amounts to a factor of $2$ multiplying one of the exponentials in~(\ref{eq:leadingOrder}): this does not contribute in the thermodynamic limit.
Noticing that, for fixed $J$,
\begin{align}\label{eq:thermodynamicLimit}
\lim_{N \rightarrow \infty} \frac{1}{N} \log \sum_{ \text{s.p.} } e^{-N \bar{S}_P(\Gamma)} = - \min_\Gamma \bar{S}_P(\Gamma),
\end{align}
where we defined the intensive quantities $\bar{S}_P =  S_P/N$, we obtain
\begin{align}\label{eq:plateauEnergies}
\epsilon_G(\Gamma) \approx \min_\Gamma ( \bar{S}_P/\tau_f ) =
- \max_\Gamma \left( \frac{J^2+\Gamma^2}{4J} ,  \frac{\Gamma}{2}    \right),
\end{align}
where the first solution is only valid for $\Gamma<J$, as discussed. Consistently with the findings of Section~\ref{sec:findingSP}, the LO result~(\ref{eq:plateauEnergies}) is equal to the result of the mean field approximation.
The energy density of the true, entangled ground state is then retrieved upon including higher order terms in the expansion of (\ref{eq:modifiedLoschmidtAmplitude}). 

\subsection{Higher Order Corrections and Quantum Phase Transitions}\label{sec:fullExpansion}
Having obtained the LO mean field approximation, we now wish to compute corrections beyond mean field. 
In the presence of multiple saddle points, it is customary to integrate Gaussian fluctuations about each saddle point and add up the relative contributions~\cite{mathematicalMethods}. Here, we assume that the expansions about different saddle points can be separately carried out and added up also for corrections beyond Gaussian. Let us discuss the conditions under which this procedure may be justified. 
Consider an integral whose integrand has several saddle points. The expansion about each SP can be seen as a way of grouping contributions together: by expanding to higher and higher order, one progressively includes trajectories further and further from each SP. 
Adding up separate expansions around different SPs is then justified provided that there is no ``overlap": the trajectories included in one expansion are not significantly contributing to any of the others. A toy example showing this is provided in Appendix~\ref{app:saddlePoint}.
In the present case, the requirement that there is no overlap is indeed satisfied: in the thermodynamic limit, Eq.~(\ref{eq:thermodynamicLimit}) implies that only one expansion contributes for each value of $\Gamma$, and no double-counting can occur.
Since the present discussion is not based on any model-specific assumption, the parameter $\Gamma$ can here be understood as a general set of parameters specifying a given Hamiltonian.
Additionally, in order to obtain finite results, each expansion should only be considered in the region of parameter space where it is convergent. 
This requirement can be physically understood as accounting for the breakdown of e.g. a large-coupling expansion in the small-coupling regime. With these caveats, let us carry out the full expansions as discussed; one has
\begin{align}\label{eq:fullExpansion}
\epsilon_G = \lim_{\tau_f\rightarrow \infty} \lim_{N\rightarrow \infty} \frac{1}{\tau_f N} \log  \sum_{ \text{s.p.} } e^{-N \bar{S}^\prime_P}
\end{align}
where, for each saddle point, the quantity $\bar{S}_P^\prime$ includes all contributions from higher order terms. Since the ground state energy density is finite and intensive, we expect $\bar{S}^\prime_P$ to be independent of the system size. By Eq.~(\ref{eq:thermodynamicLimit}), this means that
\begin{align}\label{eq:gsEnergyAssumption}
\epsilon_G = \min_\Gamma ( \bar{S}^\prime_P/\tau_f ).
\end{align}
Thus, the ground state energy is given by whichever of the series $\bar{S}^\prime_P$, obtained by expanding around the different saddle points, gives the lowest value of $\epsilon_G$ for a given value of the physical parameters. 
For example, for a quantum Ising model with fixed $J$, Eq.~(\ref{eq:gsEnergyAssumption}) implies that only one of the summands in~(\ref{eq:fullExpansion}) contributes for each choice of $\Gamma$.
The above structure then suggests an interpretation of quantum phase transitions in terms of an abrupt change in which of the series $\bar{S}^\prime_P$ yields the dominant contribution to the ground state energy. 
Let us denote by $\epsilon_\pm(\Gamma)$ the expansions corresponding to the two smallest $\bar{S}_P^\prime$ in a given range of the parameter $\Gamma$. 
In the thermodynamic limit, due to the minimum function in~(\ref{eq:gsEnergyAssumption}), the functional form of $\epsilon_G$ changes abruptly at the value $\Gamma_c$ such that $\epsilon_- < \epsilon_+$ if $\Gamma < \Gamma_c$ and $\epsilon_- > \epsilon_+$ if $\Gamma > \Gamma_c$, so that the value $\Gamma_c$ can be identified as the quantum critical point.
We refer to this condition as \textit{crossing} of the two series, although both may diverge at the critical point $\Gamma_c$ itself; see Appendix~\ref{app:saddlePoint}.
Eq.~(\ref{eq:fullExpansion}) thus implies that the GS energy density can only be non-analytic in the thermodynamic limit, and only at the point where two different series cross in the above-defined sense. This expected result is here retrieved purely on the basis of the disentanglement formalism, where it emerges naturally as a consequence of Eqs~(\ref{eq:gsEnergyFromMF_modifiedLosch}), (\ref{eq:thermodynamicLimit}) and (\ref{eq:fullExpansion}).
In the next Section, we will show that for the Ising model the expansions around the small-$\Gamma$ and large-$\Gamma$ saddle points give rise to series in $\Gamma/J$ and $J/\Gamma$ respectively. These can be identified as the small-$\Gamma$ and large-$\Gamma$ perturbative expansions of the ground state energy. 
To benchmark our analytical approach, we shall exploit the exact solvability of the Ising chain, whose ground state energy density is given by~\cite{Pfeuty1970}
\begin{align}\label{eq:GSenergyPfeuty} 
\epsilon_G = -\frac{2 \Gamma +J}{2 \pi } E\left(\frac{8 J \Gamma }{(J+2 \Gamma )^2}\right),
\end{align}
in terms of the complete elliptic integral of the second kind $E$. Expanding Eq.~(\ref{eq:GSenergyPfeuty}) for small or large $\Gamma$, one finds
\begin{align}
\epsilon_G = \begin{cases}
-\frac{J}{4} -\frac{\Gamma^2}{4 J}-\frac{\Gamma ^4}{16 J^3}-\frac{\Gamma ^6}{16 J^5} - \dots,\\ 
-\frac{\Gamma }{2}-\frac{J^2}{32 \Gamma } -\frac{J^4}{2048 \Gamma ^3} -\frac{J^6}{32768 \Gamma ^5} - \dots ,
\end{cases}\label{eq:exactPerturbativeSeries}
\end{align}
respectively. When all terms are resummed, the above perturbative series do indeed cross only at the critical point $\Gamma_c=J/2$; see Appendix~\ref{app:saddlePoint}. The small-$\Gamma$ expansion is seen to be divergent for $\Gamma>\Gamma_c$; therefore, the relative terms will not contribute in this regime. Similarly, the large-$\Gamma$ series does not contribute when $\Gamma<\Gamma_c$. Below, we will show how small- and large-$\Gamma$ expansions for $\epsilon_G$ are obtained from the disentanglement approach. 

\subsection{Feynman Rules}\label{sec:feynmanRules}
\begin{figure}
\centering
     \begin{subfigure}[b]{0.41\linewidth}
         \includegraphics[height=3.7cm]{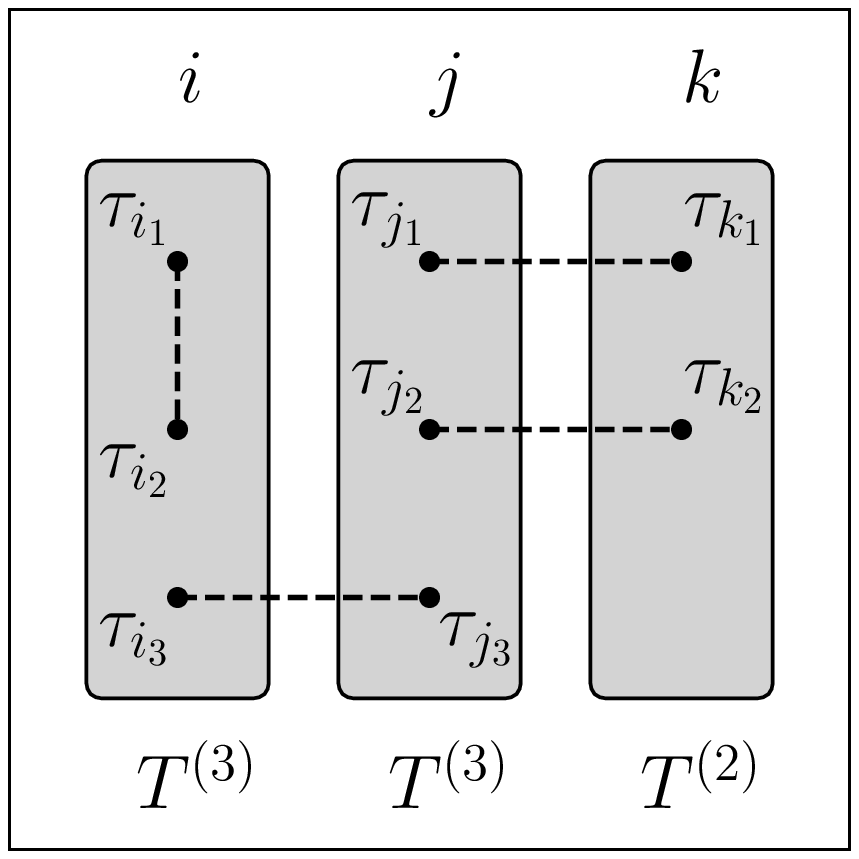}
         \caption{}
         \label{fig:vanishing_selfInteraction}
     \end{subfigure}
     \hfill
     \begin{subfigure}[b]{0.57\linewidth}
         \centering
         \includegraphics[height=3.7cm]{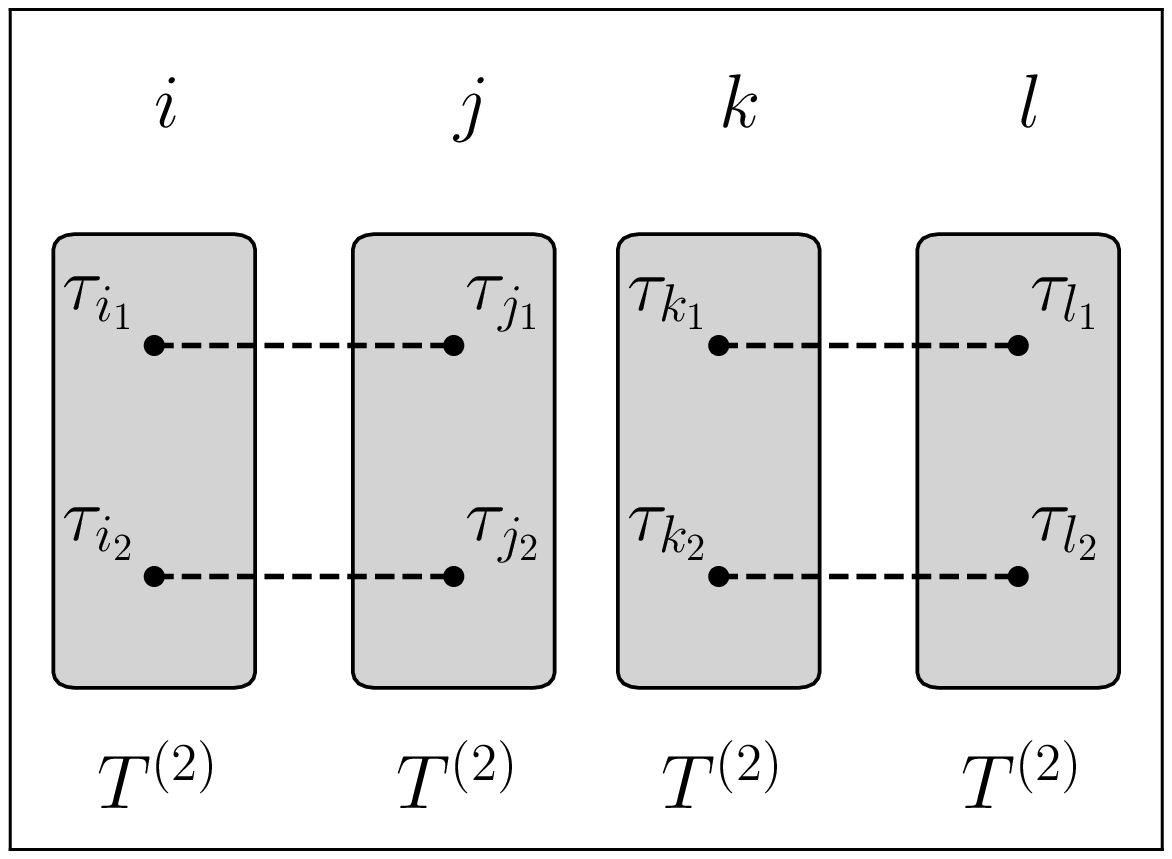}
         \caption{}
         \label{fig:vanishing_disconnected}
     \end{subfigure}\\
          \begin{subfigure}[b]{0.49\linewidth}
         \includegraphics[height=3.7cm]{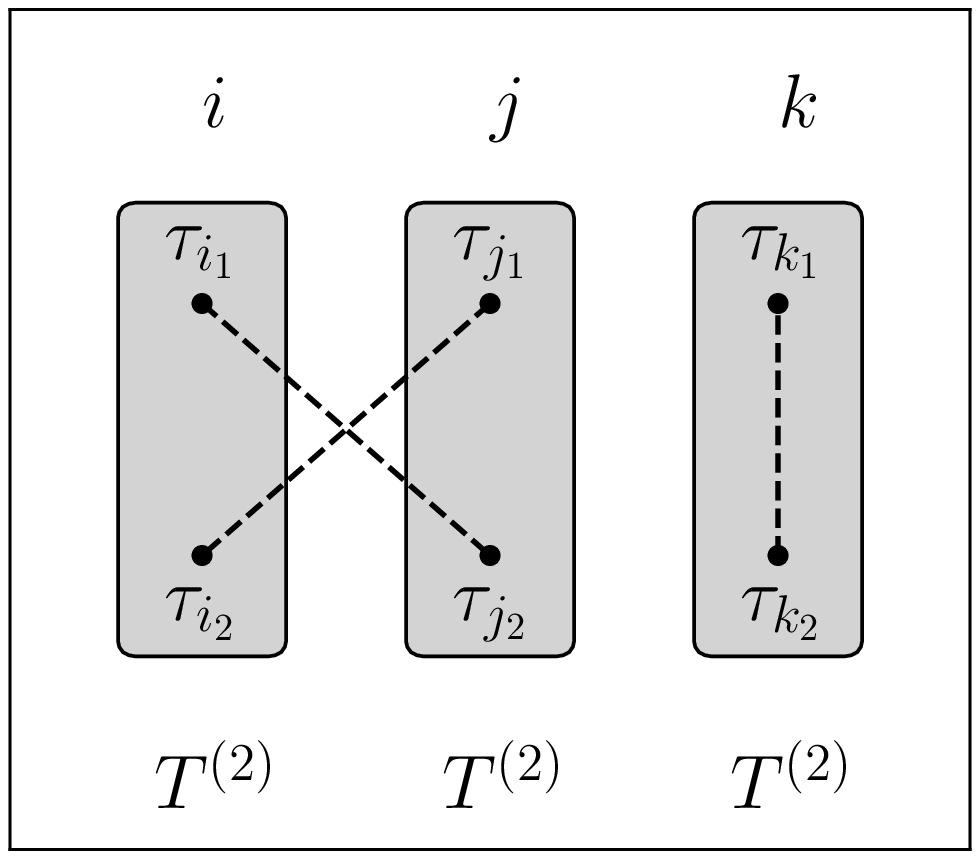}
         \caption{}
         \label{fig:vanishing_T2_selfInteraction}
     \end{subfigure}
     \hfill
     \begin{subfigure}[b]{0.49\linewidth}
         \centering
         \includegraphics[height=3.7cm]{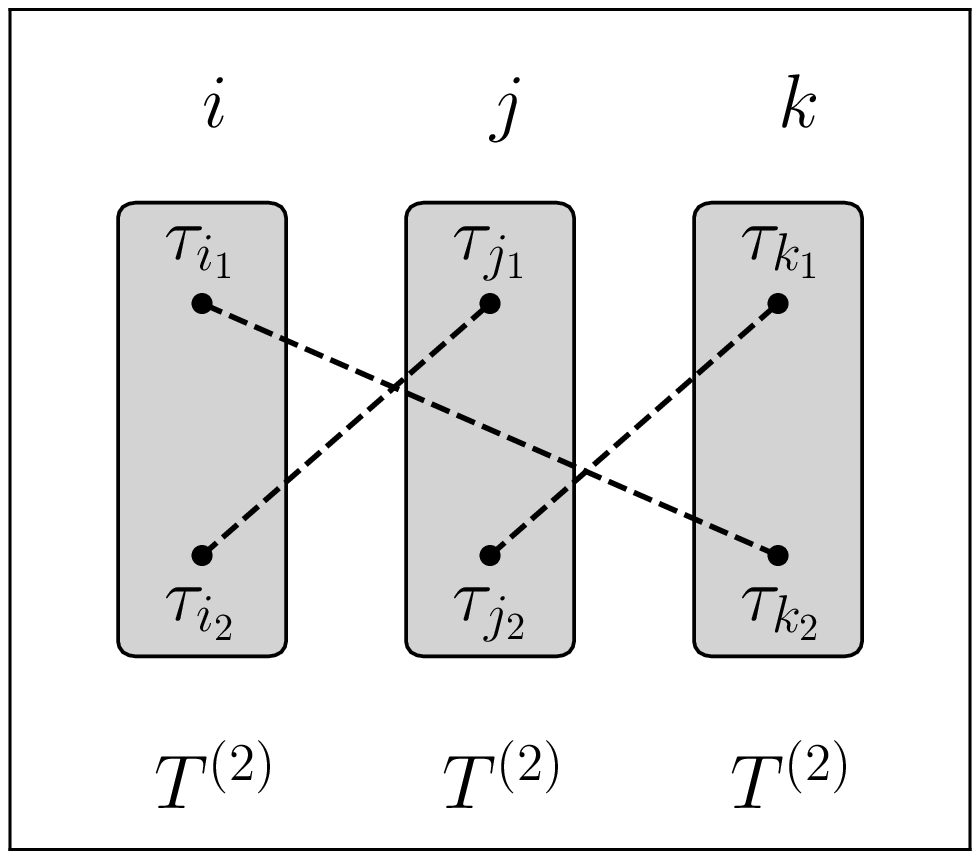}
         \caption{}
         \label{fig:vanishing_T2_odd}
     \end{subfigure}
  \caption{Examples of Feynman diagrams that give vanishing contributions to the ground state energy of the quantum Ising chain. Each square represents a vertex, as indicated underneath. Vertices are labeled by an overhead site index, e.g. $i$. Each point within a vertex also has a unique time label. Dashed lines represent propagators. (a) Diagrams where at least one line is internal to a vertex represent self-interactions and vanish identically. (b) Disconnected diagrams, where vertices that are connected by lines form disjointed clusters, do not contribute to the ground state energy. (c-d) Diagrams arising from an odd number $m$ of $T^{(2)}$ vertices always vanish (here we show $m=3$). They either (c) feature internal lines, or (d) are proportional to $\Tr \mathcal{J}^m$, where $\mathcal{J}$ is the interaction matrix; all such quantities vanishes for the Ising model. }
\label{fig:vanishingDiagrams}
\end{figure}
The evaluation of corrections to the GS energy beyond LO according to Eq.~(\ref{eq:fullExpansion}) is carried out by expanding the functional integral representation of~(\ref{eq:modifiedLoschmidtAmplitude}) around the saddle points and applying Wick's theorem. This gives rise to a set of Feynman rules, which we illustrate in this Section by considering the quantum Ising chain.
Since the expansions around the different saddle points take the same functional form, here we work in full generality, specializing only the final results to each SP.
In order to reveal the dependence of each term on the physical parameters $\Gamma$ and $J$, it is convenient to introduce dimensionless times $\btau= \gamma_P \tau$, where $\gamma_P$ will eventually be set to the appropriate plateau value~(\ref{eq:Psolutions}) for each saddle point. 
The noise action becomes
\begin{align}\label{eq:noiseActionDimensionless1}
S_0 \equiv \frac{J}{4\gamma_P} \int^{\btau_f}_{0} \mathrm{d}\tau   \sum_{ij} (\mathcal{J}^{-1})_{ij}\varphi_i(\tau) \varphi_j(\tau)
\end{align}
and the effective Loschmidt action is given by
\begin{align}\label{eq:loschmidtActionDimensionless}
S = S_0 -\int_0^{\btau_f} \mathrm{d}\btau \left[\frac{\Gamma}{2 \gamma_P}  \sum_i \xi^+_i(\btau) +\frac{J}{\gamma_P} \sum_i \varphi_i(\btau) \right].
\end{align}
It is convenient to separately consider the variations of~(\ref{eq:loschmidtActionDimensionless}) which involve functional derivatives of $\xi^+_i$ and the term originating from the noise action $S_0$: as we shall see, the latter provides a simple and physically appealing propagator\footnote{This is somewhat different from the standard QFT approach \cite{peskin1995}, where the propagator is obtained from the term in the action that is quadratic in the fields. In the present case, this procedure would not yield a propagator in closed form. Instead, it is convenient to obtain the propagator from $S_0$ and treat the remaining term of quadratic order on an equal footing to higher order terms, evaluating their contributions from Wick's theorem.}. 
Thus, for each SP we expand the action~(\ref{eq:loschmidtActionDimensionless}) as
\begin{align}\label{eq:expandedAction}
S &=  S_P + S_0 - \sum_{n=2}^\infty T^{(n)}
\end{align}
where we defined
\begin{align}
\begin{split}
 T^{(n)}  & \equiv  \\ 
& \hspace{-6mm}\frac{1}{n!} \sum_i \int_0^{\btau_f}\hspace{-2mm}\dots\int_0^{\btau_f}\hspace{-2mm} S^{(n)}(\btau_1,\dots,\btau_n) \varphi_{i} (\btau_1) \dots \varphi_{i}(\btau_n) \dd \btau_1 \dots   \dd \btau_n , 
 \end{split}
 \end{align}
 \begin{align}
 \begin{split}
S^{(n)}(\btau_1,\dots,\btau_n)   \equiv  - \frac{\Gamma}{2 \gamma_P} \int_0^{\btau_f} \dd\btau \frac{\delta^n \xi^+_i(\btau) }{\delta \varphi_{i} (\btau_1)\dots \delta\varphi_{i}(\btau_n)}\Big\vert_P ,
 \end{split}
\end{align}
and we exploited translational invariance and $\delta \xi^+_i/\delta\varphi_j \propto \delta_{ij}$. The functional integral (\ref{eq:loschmidtAmplitude}) can the be expanded as
\begin{align}
A(\tau_f) = \sum_{ \text{s.p.} } e^{-S_{P}}  \Big\langle \sum_{m=0}^\infty \frac{(\sum_{n=2}^\infty T^{(n)} )^m}{m!} \Big\rangle_0 , \label{eq:expansion}
\end{align}
where the notation $\langle \dots \rangle_0$ denotes averaging with respect to the noise action~(\ref{eq:noiseActionDimensionless1}).
Each term in Eq.~(\ref{eq:expansion}) can be evaluated using Wick's theorem. The propagator $\Delta$, which accounts for interactions in the system, can be read off from the quadratic action~(\ref{eq:noiseActionDimensionless1}) and is found to be proportional to the interaction matrix: 
\begin{align}\label{eq:propagator}
\Delta_{ij}(\btau,\btau^\prime) = 2\frac{\gamma_P}{J} \mathcal{J}_{ij} \delta(\btau-\btau^\prime).
\end{align}
The series obtained from~(\ref{eq:expansion}) can then be formally re-exponentiated, giving Eq.~(\ref{eq:fullExpansion}).
Consider the averages $\langle \dots \rangle_0$ in Eq.~(\ref{eq:expansion}). We define the \textit{order} of a term $\langle T^{(n_1)} \dots T^{(n_m)} \rangle_0$ to be $l=\sum_{j=1}^m n_j$. 
Wick's theorem implies that terms of odd order vanish identically, while terms of even order are obtained by summing over all the possible replacements of pairs of fields $\phi_i(\btau_i)$, $\phi_j(\btau_j)$ by propagators $\Delta_{ij}(\btau_i,\btau_j)$. 
The evaluation of a given term in~(\ref{eq:expansion}) is simplified by means of a diagrammatic representation and the associated Feynman rules:
\begin{itemize}
\item Each $T^{(n)}$ provides a \textit{vertex} and contributes a factor $S^{(n)}/n!$. Diagrammatically, a vertex is represented as $n$ points arranged inside a box. Each vertex is labeled by a site index, e.g. $j$. Individual points belonging to a given vertex are additionally distinguished by a unique time label, $\btau_{j_1} \dots \btau_{j_n}$. 
\item One then sums over all possible ways of joining pairs of points by lines; a line joining the points labeled by $(j , \btau_{j})$, and  $(k , \btau_{k})$ gives a propagator $\Delta_{jk}(\btau_{j},\btau_{k})$.
\item The resulting quantity is then integrated over all times $\btau_i$; the integrals run between $0$ and $\btau_f$.
\item Finally, all site indices are summed over.
\end{itemize} 
Examples of this diagrammatic representation are given in Figs~\ref{fig:vanishingDiagrams}, \ref{fig:T4} and \ref{fig:T6}, discussed below.
In more usual field theories, such as $\phi^4$, vertices are typically represented by single points \cite{peskin1995}; in the above rules, this would correspond to setting all $\btau_{j_i}$ to the same value. The fact that here vertices consist of separate points is due to the non-locality in time of the action~(\ref{eq:expandedAction}).
To simplify the evaluation of higher order terms, we identify two classes of diagrams which do not contribute. The first class includes diagrams where lines join points within the same vertex: these are \textit{self-interaction} diagrams. An example, originating from $\langle T^{(3)}T^{(3)}T^{(2)} \rangle_0$, is shown in Fig.~\ref{fig:vanishing_selfInteraction}. Such diagrams feature at least one term of the form $\Delta_{ii} \propto \mathcal{J}_{ii}$, which is identically zero for the Hamiltonian~(\ref{eq:Ising}) at hand.
The second class of non-contributing terms includes \textit{disconnected diagrams}, in which the vertices joined by internal lines form disjointed clusters. This class includes the diagram in Fig.~\ref{fig:vanishing_disconnected}, which is produced by $\langle T^{(4)}T^{(4)} \rangle_0$.
It is easy to see that, due to the Feynman rules and the form of the propagator~(\ref{eq:propagator}), a connected cluster of vertices gives a contribution proportional to the system size $N$. A term with $m$ disconnected clusters of vertices is then proportional to $N^m$.
The origin of these terms can be understood by considering the expansion
\begin{align}\label{eq:disconnectedTerms}
e^{-N\bar{S}^\prime} = 1 - N \bar{S}^\prime + \frac{1}{2!} N^2 (\bar{S}^\prime)^2 +\dots \,  .
\end{align}
The terms $\langle \dots \rangle_0$ in~(\ref{eq:expansion}), obtained from Wick's theorem, correspond to the right-hand side of~(\ref{eq:disconnectedTerms}). The ground state energy must be intensive and finite in the thermodynamic limit; this implies that terms proportional to higher powers of $N$ must cancel out when exponentiating the series in~(\ref{eq:expansion}) to obtain Eq.~(\ref{eq:fullExpansion}). This is precisely what happens in Eq.~(\ref{eq:disconnectedTerms}). Since the desideratum here is $\bar{S}^\prime$, we only need to consider terms proportional to $N$: these are given by connected diagrams. 
Finally, we note that for the quantum Ising model all diagrams with an odd number $m$ of vertices $T^{(2)}$ vanish: any such diagram is either self-interacting (Fig.~\ref{fig:vanishing_T2_selfInteraction}), or it gives rise to a term $\propto \Tr \mathcal{J}^m$, which vanishes for $\mathcal{J}_{ij} =(\delta_{i, j+1} + \delta_{i+1,j})/2$ (Fig.~\ref{fig:vanishing_T2_odd}).

In Section~\ref{sec:higherOrder}, we apply the Feynman rules derived in this Section to compute higher order corrections to the ground state energy of the quantum Ising chain.
Before turning to this explicit example, in the next Section we complete our theoretical overview by considering how the terms in Eq.~(\ref{eq:expansion}) depend on the physical parameters of the model, elucidating the relation between the expansion about the saddle points and perturbation theory.\\

\subsection{Relation to Perturbation Theory}\label{sec:perturbationTheory}
In order to understand the nature of the terms produced by the expansion~(\ref{eq:expansion}) we need to consider the higher variations of the action, $S^{(n)}$ with $n\geq 2$. It is convenient to compute these variations by initially imposing a time ordering $\btau_n> \dots >\btau_1$, and then symmetrizing the result with respect to the times $\btau_i$. With said ordering, one obtains for the second variation
\begin{widetext}
\begin{align}\label{eq:secondVariation}
S^{(2)}(\btau_1,\btau_2)= - \left(\frac{J}{\gamma_P}\right)^2 \int_0^{\btau_f}  \frac{\Gamma}{\gamma_P} \overline{\Xi}_i(\bs_1,\btau_1) \left[ \theta(\bs_1- \btau_2)  - \frac{\Gamma}{\gamma_P}\int_{\btau_2}^{\bs_1} \overline{\Xi}_i(\bs_2,\btau_2) \mathrm{d}\bs_2 \right] \dd\bs_1 \Big\vert_P ,
\end{align}
\end{widetext}
where we defined $\Xi(\bs,\btau_1) \equiv\frac{J}{\gamma_P} \overline{\Xi}(\bs,\btau_1)$ to make the dependence on physical parameters manifest. Eq.~(\ref{eq:secondVariation}) shows that all variations $S^{(n)}$ with $n\geq 2$ can be expressed in terms of integrals of the first variation $\Xi_i$. Schematically, one obtains $S^{(n+1)}$ from $S^{(n)}$ by summing over all possible ways of replacing $$\overline{\Xi} \rightarrow \theta  -\frac{\Gamma}{\gamma_P} \int \overline{\Xi}  $$ and multiplying by $J/ \gamma_P$. When evaluated at the SP, each $\overline{\Xi}$ gives a factor of $\xi_P$ and an exponential depending on the dimensionless times $\bar{\tau}_i$ only. Therefore, the $n$-th variation (with $n \geq 2$) evaluated at the plateau must be of the form
\begin{align}\label{eq:structureHigherVariations}
S^{(n)}(\btau_1, \dots , \btau_n) = \left(\frac{J}{\gamma_P}\right)^{n} \sum_{m=1}^{n} C_{n,m}(\btau) \left(\frac{\Gamma \xi^+_P}{\gamma_P} \right)^m  ,
\end{align}
where $C_{n,m}(\btau)$ are dimensionless functions depending only on the $\btau_i$, which do not involve any factor of $\Gamma$, $J$ or $\gamma_P$. Using the form~(\ref{eq:structureHigherVariations}) of higher variations, it is possible to determine the structure of the terms in Eq.~(\ref{eq:expansion}).
Terms of odd order $l=2m+1$ do not contribute due to Wick's theorem; see the discussion in Section~\ref{sec:feynmanRules}.
Any given term of even order $l=2m$ features a product of variations, whose orders add up to $2m$, and $m$ propagators, each of which carries a factor $\gamma_P/J$.
Bringing everything together and substituting the SP values~(\ref{eq:Psolutions}), a general $2m-$th order term $\mathcal{T}^{2m}$ can be written as
\begin{align}\label{eq:structureTerms}
\mathcal{T}^{2m}= \begin{cases}
\sum_{n=1}^\infty \bar{C}_{n,2m} \left(\frac{\Gamma}{J}\right)^{2n} \\
\bar{D}_{2m} \left(\frac{J}{\Gamma}\right)^m 
\end{cases}
\end{align}
where the top and bottom solutions refer to the small-$\Gamma$ and large-$\Gamma$ expansion respectively, and $\bar{C}_{n,2m}$ and $\bar{D}_{2m}$ are dimensionless constants which do not depend on $J$ or $\Gamma$. The series of even powers of $(\Gamma/J)$ in the former case arises from Taylor expanding $\Gamma \xi^+_P/\gamma_P$. Eq.~(\ref{eq:structureTerms}) thus shows that the expansions around the saddle points give rise to series in $(\Gamma/J)^2$ and $J/\Gamma$. 
These can be identified with the perturbative series as follows. Eq.~(\ref{eq:fullExpansion}) is valid for any value of $\Gamma$ and, due to the thermodynamic relation~(\ref{eq:gsEnergyAssumption}), only one expansion at a time contributes. Consider the $\Gamma \rightarrow 0$ limit; in this case, the LO term~(\ref{eq:plateauAction}) of the small-$\Gamma$ expansion gives the exact value of the ground state energy. 
For finite but sufficiently small $\Gamma/J \ll 1$, the small-$\Gamma$ expansion will still be the dominant one and give the ground state energy $\epsilon_G$. The small-$\Gamma$ expansion must therefore be equal to the $\Gamma/J$ perturbative series for $\epsilon_G$, as they are both series in $\Gamma/J$ and they both add up to $\epsilon_G$. A symmetric argument holds for the large-$\Gamma$ series in the corresponding limit. 
Although we illustrated this result for the Ising model, the correspondence is expected be more generally valid: both expansions yield the exact free energy $\epsilon(\Gamma)$ in the strong-coupling $(\Gamma=0)$ or strong-field $(\Gamma\rightarrow\infty)$ limits, and by smoothness of $\epsilon(\Gamma)$ it can be expected that these expansions will continue to give the same result provided no phase transition is crossed. In the case of the Ising model, the argument applies either side of the transition, as discussed above. 

Eq.~(\ref{eq:structureTerms}) further shows that the large-$\Gamma$ expansion (bottom case) is in order-by-order correspondence to the perturbative series: terms of order $2m$ are proportional to $(\Gamma/J)^m$. 
On the other hand, the small-$\Gamma$ expansion (top case of Eq.~(\ref{eq:structureTerms})) is not in one-to-one correspondence to perturbation theory: one needs to sum over $m$ in order to retrieve the perturbative series in $\Gamma/J$, since each of the terms in Eq.~(\ref{eq:structureTerms}) may in principle contain various powers of $\Gamma/J$. Such different behavior of the two expansions is due to the nature of the plateau configuration or, equivalently, the MF ground state. For large $\Gamma$, this is just the $\Gamma=\infty$ ground state $\ket{\Rightarrow}$; the large-$\Gamma$ expansion is thus equivalent order-by-order to the perturbative series around $\Gamma=\infty$. On the other hand, for small $\Gamma$ the MF ground state is not simply given by the $\Gamma=0$ ground state $\ket{\Downarrow}$. Consider for instance the MF magnetization, given in Appendix~\ref{app:meanField}; this can be expanded as a Taylor series featuring all even powers of $(\Gamma/J)$. An expansion around the MF ground state is therefore not expected to be in order-to-order correspondence with a perturbative expansion around $\Gamma=0$.

One more comment is due concerning even and odd powers in the two expansions. From expanding the exact ground state energy of the quantum Ising chain as in~(\ref{eq:exactPerturbativeSeries}), we see that the perturbative expression for $\epsilon_{G}/J$ around $\Gamma=0$ features only even powers of $\Gamma/J$ and, similarly, the perturbative expansion of  $\epsilon_{G}/\Gamma$ around $\Gamma=\infty$ contains only even powers of $J/\Gamma$.
This result is immediately retrieved from Eq.~(\ref{eq:structureTerms}) for the small-$\Gamma$ expansion, and is due to spontaneous symmetry breaking at the MF level. However, odd powers of $J/\Gamma$ are not excluded a priori in the large-$\Gamma$ expansion. The necessary cancellation must therefore originate from the vanishing of the $\bar{D}_{2m}$ coefficient in~(\ref{eq:structureTerms}) when $m$ is odd. We explicitly show an example of such cancellation when computing higher-order corrections to $\epsilon_G$ in Section~\ref{sec:higherOrder}. 

We have thus determined the structure of the terms produced by expanding about the saddle points, and clarified the relation of such expansions to perturbation theory. In summary, the small-$\Gamma$ and large-$\Gamma$ expansions, taken as a whole, are respectively equal to the full perturbative series around $\Gamma=0$ and $\Gamma=\infty$. This correspondence is satisfied order-by-order for the large-$\Gamma$ expansion, but only when resumming the whole series for the small-$\Gamma$ expansion.
This completes an overview of the field theoretical approach; in the next Section, we apply the concepts discussed so far to compute corrections to the ground state energy of the quantum Ising chain.

\subsection{Example: NLO and NNLO Corrections to the Ground State Energy}\label{sec:higherOrder}
\begin{figure}
\centering
          \begin{subfigure}[b]{0.49\linewidth}
         \includegraphics[height=4.1cm]{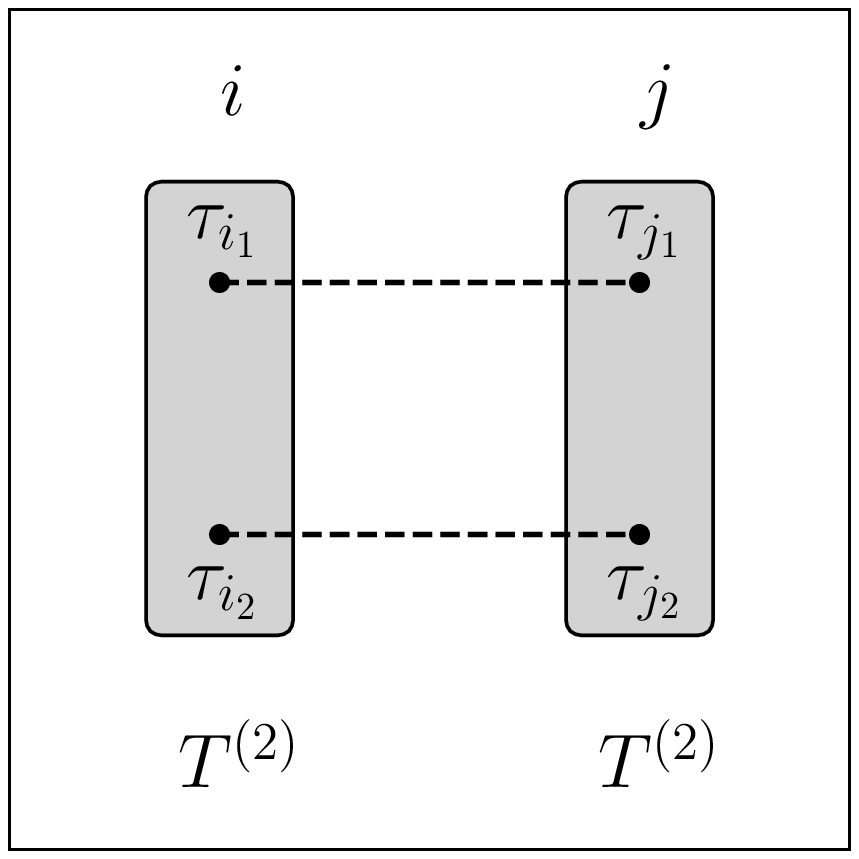}
         \caption{}
         \label{fig:T4par}
     \end{subfigure}
     \hfill
     \begin{subfigure}[b]{0.49\linewidth}
         \centering
         \includegraphics[height=4.1cm]{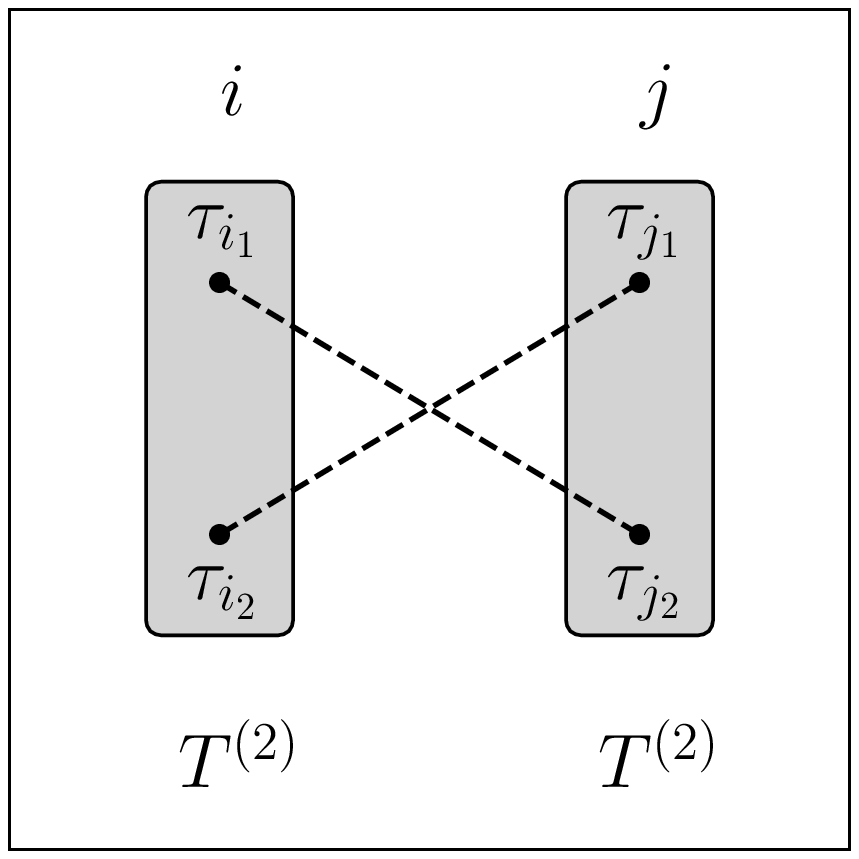}
         \caption{}
         \label{fig:T4cross}
     \end{subfigure}
  \caption{Diagrams contributing to $\mathcal{T}^{(4)}$.}
\label{fig:T4}
\end{figure}

\begin{figure}
\centering
          \begin{subfigure}[b]{0.49\linewidth}
         \includegraphics[height=4.1cm]{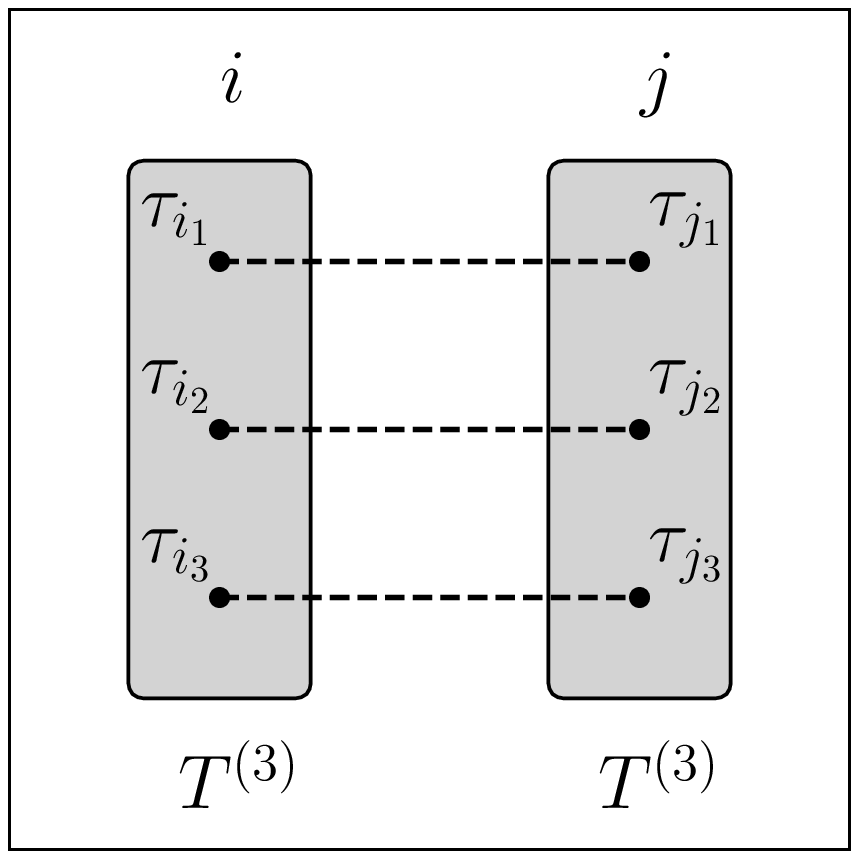}
         \caption{}
         \label{fig:T6par}
     \end{subfigure}
     \hfill
     \begin{subfigure}[b]{0.49\linewidth}
         \centering
         \includegraphics[height=4.1cm]{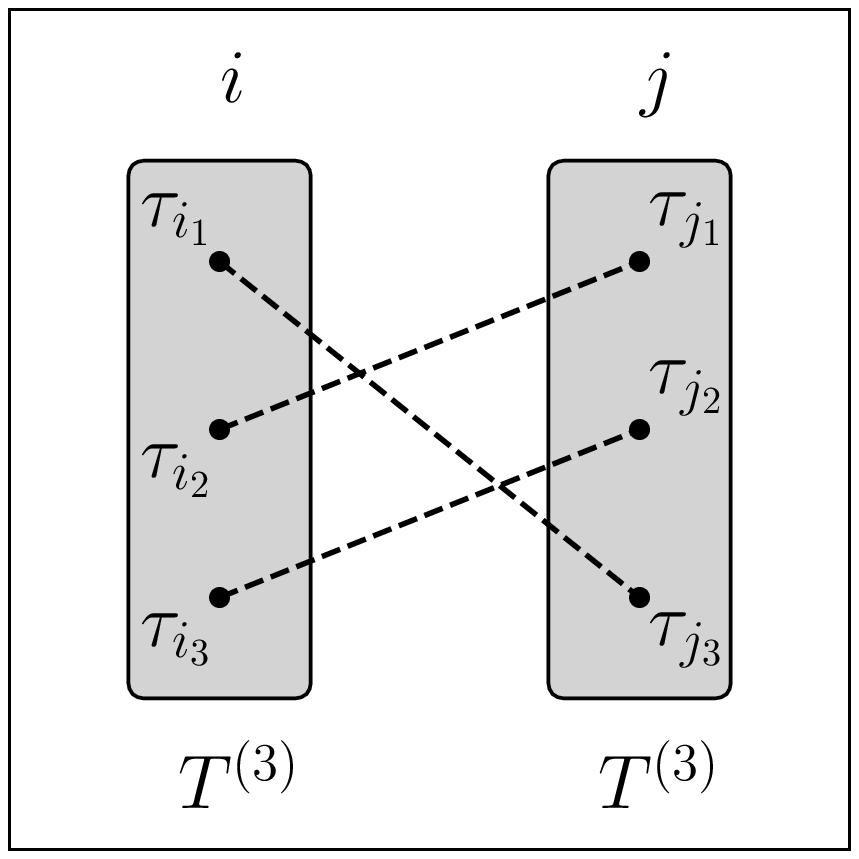}
         \caption{}
         \label{fig:T6cross}
     \end{subfigure}
  \caption{Diagrams contributing to $\mathcal{T}^{(6)}$.}
\label{fig:T6}
\end{figure}

\begin{figure}
\centering
  \includegraphics[width=\linewidth, trim=0 0 0 30, clip]{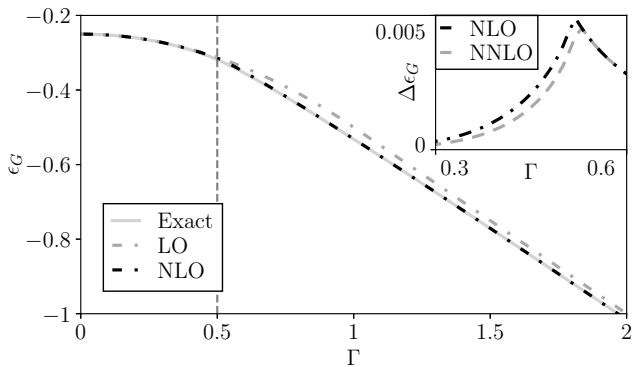}
  \caption{Approximations to the ground state energy of the quantum Ising chain in the thermodynamic limit. Main panel: comparison of the exact energy to the approximations obtained by expanding about the SP to leading order (LO), corresponding to the mean field solution, and to next-to-leading order (NLO), which includes corrections beyond mean field. Inset: difference $\Delta \epsilon_G$ between the exact solution and the NLO (black dash-dotted line) and next-to-next-to-leading order (NNLO, gray dashed line) results. Increasing the order of the expansion monotonically leads to a better approximation.}
\label{fig:higherOrderCorrections}
\end{figure}

In order to illustrate the machinery introduced in the previous Sections, we compute the next-to-leading order (NLO) and next-to-next-to-leading order (NNLO) corrections to the ground state energy density of the quantum Ising chain.
The lowest order correction is naively given by $\langle T^{(2)} \rangle_0$; this term however vanishes, since the corresponding diagram is self-interacting: see the discussion in Section~\ref{sec:feynmanRules}. The NLO correction is then given by the next-higher term, which is of order four:
\begin{align}\label{eq:orderFourCorrection}
\mathcal{T}^{(4)}= \frac{1}{2}\langle T^{(2)} T^{(2)} \rangle_0 + \langle T^{(4)} \rangle_0 .
\end{align}
The second term on the right hand side of Eq.~(\ref{eq:orderFourCorrection}) vanishes similarly to $\langle T^{(2)} \rangle_0$; the remaining term corresponds to the diagrams in Fig.~\ref{fig:T4}, and gives
\begin{align}\label{eq:NLOterm}
\frac{1}{2}\langle T^{(2)} T^{(2)}\rangle_0 = \btau_f N\begin{cases} \frac{\Gamma^4}{32 J^4} \\
 \frac{J^2}{32 \Gamma^2}
\end{cases}
\end{align}
where the top and bottom results are obtained from the small-$\Gamma$ and large-$\Gamma$ SPs respectively.
Thus, the NLO approximation to the ground state energy to  is given by
\begin{align}\label{eq:NLO}
\epsilon_G \approx - \max_\Gamma \left(   \frac{J}{4}+ \frac{\Gamma^2}{4J}  + \frac{\Gamma^4}{32J^3} , \frac{\Gamma}{2} + \frac{J^2}{32\Gamma}  \right) .
\end{align}
As discussed in Section~\ref{sec:fullExpansion}, each of the two series in Eq.~(\ref{eq:NLO}) can only be considered within its radius of convergence. 
In practice, one typically does not have access to the full series; the radius of convergence can then be estimated by imposing that each term be smaller than the lower-order one. 
In the present case, this criterion indicates that the small-$\Gamma$ series is valid for $\Gamma<1$, while the large-$\Gamma$ series is valid for $\Gamma>1/4$. Including the NLO correction as discussed provides an improvement over the LO approximation for all values of $\Gamma$; see Fig.~\ref{fig:higherOrderCorrections}.
Consistently with the discussion in Section~\ref{sec:perturbationTheory}, Eq.~(\ref{eq:NLO}) matches the result of second-order perturbation theory about $\Gamma=\infty$, which is in one-to-one correspondence with the expansion around the large-$\Gamma$ SP. On the other hand, in order to match the term of order $\Gamma^4/J^3$ of the perturbative series in~(\ref{eq:exactPerturbativeSeries}), one needs to include higher order contributions from the small-$\Gamma$ SP expansion: this is again consistent with the earlier discussion. 
The next-higher correction to the ground state energy, NNLO, is of order $6$, and is given by
\begin{align}\label{eq:orderSixCorrection}
\mathcal{T}^{(6)}= \frac{1}{3!}\langle T^{(2)}T^{(2)} T^{(2)} \rangle_0 + \frac{1}{2!} \langle T^{(3)}T^{(3)} \rangle_0 .
\end{align}
The first term in Eq.~(\ref{eq:orderSixCorrection}) vanishes because it features an odd number of $T^{(2)}$ vertices; see the discussion in Section~\ref{sec:feynmanRules} and in particular Fig.~\ref{fig:vanishingDiagrams}(c-d). The non-vanishing diagrams are shown in Fig.~\ref{fig:T6}. They can be evaluated to give
\begin{align}\label{eq:NNLOterms}
\mathcal{T}^{(6)} = \btau_f N\begin{cases} \frac{\Gamma ^4}{64 J^4}-\frac{\Gamma ^6}{64 J^6} \\
0
 \end{cases}
\end{align}
where again the top result corresponds to the small-$\Gamma$ expansion and the bottom result to the large-$\Gamma$ expansion. Eq.~(\ref{eq:NNLOterms}) shows that the NNLO correction from the large-$\Gamma$ expansion vanishes. This was anticipated in Section~\ref{sec:perturbationTheory}, and is due to the fact that the SP expansions and the perturbative series must coincide order-by-order; by Eq.~(\ref{eq:structureTerms}), the large-$\Gamma$ NNLO correction would be proportional to $J^3$, but no such term appears in the perturbative series~(\ref{eq:exactPerturbativeSeries}): the coefficient multiplying $J^3$ must therefore vanish. 
Including the NNLO term~(\ref{eq:NNLOterms}) leads to a further improvement in the approximation to the GS energy, as shown in the inset of Fig.~\ref{fig:higherOrderCorrections}.

We have thus illustrated how the disentanglement formalism provides a method to analytically approximate ground state expectation values. The leading order saddle point result corresponds to mean field, while analytical corrections beyond mean field can be systematically obtained by means of Feynman diagrams, which we explicitly showed. For simplicity, in this Section we focused on the ground state energy, but an analogous procedure can be carried out for other observables by expanding the appropriate effective action $S_\mathcal{O}$, defined as in Eq.~(\ref{eq:effectiveAction}). 
The machinery introduced above is qualitatively different from other existing diagrammatic techniques. Diagrammatic expansions for spin systems conventionally make use of a formal mapping to an auxiliary fermionic system with imaginary chemical potential \cite{popov1988}; see e.g. Refs~\cite{KulaginPRB2013,KulaginPRL2013,huang2016} for recent applications. In contrast, the present field theory is directly formulated in spin language. Furthermore, the present method does not yield a perturbative expansion about a classical or non-interacting limit, respectively given by $\Gamma=0$ and $J=0$  for the Ising model, but an expansion about the mean field solution. These are in general different, e.g. the mean field magnetization for the $D$-dimensional Ising model is given by $\sqrt{D^2 J^2-\Gamma^2}/2$ for $\Gamma<JD$, whose small$-\Gamma$ expansion features all even powers of $\Gamma$; this cannot be matched by any order of perturbation theory.

Beside the analytical field theoretical formalism outlined in this Section, the disentanglement method can alternatively be used as a numerical tool; we discuss this approach in the following Section.

\section{Importance Sampling}\label{sec:importanceSampling}
\subsection{Measure Transformation}
Knowledge of the saddle point trajectory can be used to implement an importance sampling numerical algorithm.
For this application, it is convenient to work with the diagonal form~(\ref{eq:noiseActionDiagonal}) of the noise action, involving the fields $\phi_i^a$. The corresponding SP values can be readily determined using $\phi^a_i \vert_{\text{SP}}= \sum_{bj} O^{ab}_{ij} \varphi_j^b \vert_{\text{SP}}$. The key step of the proposed approach consists in using $\phi_{\text{SP}}\equiv \{ \phi_i^a \vert_{\text{SP}} \}$ to perform the change of variables
\begin{align}\label{eq:changeOfVariablesSP}
\phi_i^a = (\phi_{\text{SP}})^a_i + \phi^{\prime a}_i
\end{align}
in the functional integral for a given observable:
\begin{align}
\langle\hat{\mathcal{O}}\rangle = e^{-S_0[\phi_{\text{SP}}]}\hspace{-1.5mm} \int \mathcal{D}\phi^\prime e^{-S_0 [\phi^\prime]} e^{-\int \mathrm{d}\tau \phi^\prime(\tau) \cdot \phi_{\text{SP}}(\tau)  } f_\mathcal{O}[\phi_{\text{SP}}+\phi^\prime] \label{eq:sampleAroundSP},
\end{align}
where $\phi^\prime \equiv \{ \phi^{\prime a}_i \}$ and $\phi_{\text{SP}} \cdot \phi^\prime \equiv \sum_{ia} (\phi_{\text{SP}})_{i}^a \phi_i^a$. Due to the Gaussianity of the noise action $S_0$, Eq.~(\ref{eq:sampleAroundSP}) can be evaluated numerically in the spirit of the stochastic approach of Refs~\cite{stochasticApproach,nonEquilibrium}; this amounts to averaging a biased function over realizations of Gaussian-distributed stochastic processes $\phi^\prime$:
\begin{align}\label{eq:biasedAverage}
\langle\hat{\mathcal{O}}\rangle =e^{-S_0[\phi_{\text{SP}}]} \langle e^{-\int \mathrm{d}\tau \phi^\prime(\tau) \cdot \phi_{\text{SP}}(\tau)  } f_\mathcal{O}[\phi_{\text{SP}}+\phi^\prime] \rangle_{\phi^\prime} .
\end{align}
By construction, the stochastic processes $\phi^\prime$ featured in~(\ref{eq:biasedAverage}) are fluctuations about the dominant SP trajectories.
In contrast, when Eq.~(\ref{eq:effectiveAction}) is sampled directly according to~(\ref{eq:noiseActionDiagonal}), trajectories close to $\phi(\tau)=0$ are sampled preferentially, even though they may give a small contribution to the integral.
Compared to more usual path integral approaches, here we \textit{do not} truncate to a given order in the fluctuations, so that Eq.~(\ref{eq:sampleAroundSP}) does not constitute a semiclassical approximation. Instead, a change of variables is used to bias the sampling towards important trajectories; the exactness of Eq.~(\ref{eq:effectiveAction}) is thus fully preserved in Eq.~(\ref{eq:sampleAroundSP}).
This change of variables can be seen as a particular measure (or Girsanov) transformation \cite{girsanov,kloeden}; in the context of stochastic processes, this constitutes the continuum version of importance sampling. 

\subsection{Numerical Results}
To illustrate our method, we apply the measure transformation approach to the $D$-dimensional quantum Ising model~(\ref{eq:Ising}) for $D\in\{1,2,3\}$ by numerically computing different observables from stochastic simulations. In our numerical simulations and in the remainder of this Section we set $J=1$. In the stochastic approach, ground state expectation values are computed according to Eq.~(\ref{eq:GSexpectationValue}). By appropriately choosing $\hat{U}_0$ in Eq.~(\ref{eq:initialState}), any initial state $\ket{\psi_0}$ can be considered. Following the discussion of Section~\ref{sec:findingSP}, it is convenient to choose the initial state to be the mean-field ground state for the desired value of $\Gamma$, $\ket{\text{MF}}$. 
As anticipated, this is equivalent to initializing the system at the plateau SP configuration, $\xi^+_i(0) = \xi^+_P$, $\xi^z_i(0) = \log ( 1+|\xi^+_P|^2)$. 
For observables computed from~(\ref{eq:GSexpectationValue}), the plateau values are fixed points of the saddle point equation, such that one may perform the change of variables~(\ref{eq:changeOfVariablesSP}) with $\phi_{\text{SP}}(\tau)= \phi_P$; see Appendix~\ref{app:saddlePoint}. 
The subsequent imaginary time evolution then projects the wavefunction from the MF to the true ground state.
The SDEs implementing the imaginary time evolution are solved using the Euler scheme \cite{kloeden}.
The statistical uncertainty on each quantity obtained as a classical average is computed by partitioning the data set into $n_B$ batches of independent simulations. We use $n_B=100$ unless otherwise stated. The data within each batch are averaged, and the standard deviation $\sigma$ of each quantity over the $n_B$ batches is then computed. The uncertainty on the mean is estimated as the standard error $\sigma/\sqrt{n_B}$. Since observables are obtained from the ratio~(\ref{eq:GSexpectationValue}), we apply the appropriate uncertainty propagation formula
\begin{align}\label{eq:errorProp}
\frac{\sigma^2(X/Y)}{(X/Y)^2} = \frac{\sigma^2(X)}{X^2}  +\frac{\sigma^2(Y)}{Y^2}   - 2\frac{ \cov(X,Y)}{X Y } ,
\end{align}
where $\cov(X,Y)$ is the covariance of $X,Y$.
It is also convenient to exploit the real-valuedness of the numerator and denominator of~(\ref{eq:GSexpectationValue}) to consider only the real parts of quantities obtained from averaging. We estimate the error on an observable $\mathcal{O}$ to be $\sigma(\mathcal{O})$ computed as above. 
 
\begin{figure}
\centering
\includegraphics[width=\linewidth]{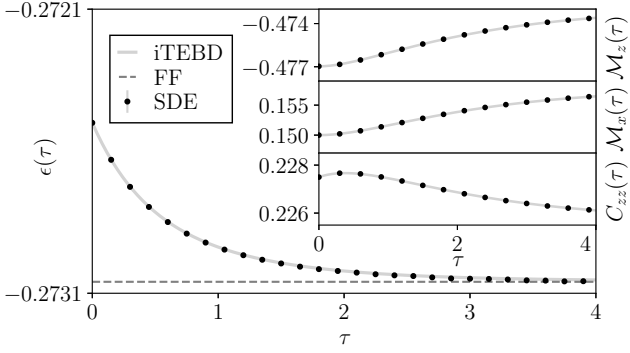}
\caption{Imaginary time evolution for the one-dimensional quantum Ising chain~(\ref{eq:Ising}). We consider the ground state energy density $\epsilon$ (main panel), and local observables defined in the main text, including the longitudinal and transverse magnetization, $\mathcal{M}_z$ and $\mathcal{M}_x$, and the nearest neighbor correlation function $C_{zz}$ (insets, top to bottom). The system is initialized in the mean field ground state for $\Gamma=0.3$ and subsequently evolved in Euclidean time towards the true ground state for the same value of $\Gamma$. We show results obtained for a system of $N=101$ spins using the importance sampling scheme discussed in the main text. We compare the results obtained from solving the SDEs for the importance sampling scheme (dots) against numerically exact imaginary time evolution in the thermodynamic limit, given by iTEBD (solid lines); we find excellent agreement. At the stopping time $\tau= 4$, the relative error compared to the exact ground state energy is of order $10^{-5}$, as found by comparison with the exact free-fermionic (FF) solution (dashed horizontal line). Our results were obtained from $10^8$ realizations of the stochastic process with time step $\Delta t=0.005$. The error bars, discussed in the main text, are not visible on the scale of the plot.}
\label{fig:1Dnumerics}
\end{figure}

\begin{figure}
\centering
  \includegraphics[width=\linewidth, trim=0 0 0 16, clip]{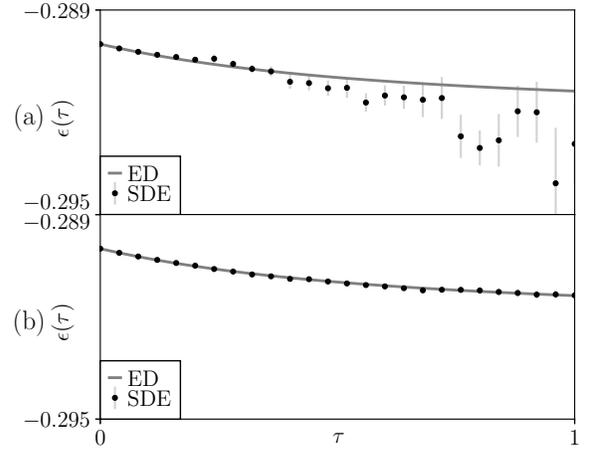}
  \caption{Comparison of direct and importance sampling for the quantum Ising chain. We consider a system with $N=15$ spins, initialized in the mean field ground state for $\Gamma=0.4$ and evolved with the same $\Gamma$ using (a) direct sampling and (b) importance sampling. We compute the Euclidean time evolution of the ground state energy from $2\times 10^4$ simulations, performed using the same time step $\Delta\tau=0.01$ for both methods; the corresponding results are compared to ED (solid line). It can be seen that the importance sampling method produces significantly better results for the same computational cost. 
Due to the comparatively small number of simulations, we do not divide the data set into batches and estimate the uncertainty by computing the standard deviation for the numerator and denominator of~(\ref{eq:GSexpectationValue}) over the full data set and applying Eq.~(\ref{eq:errorProp}).
The bars thus obtained show the much faster growth of fluctuations for direct compared to importance sampling. Both simulations took approximately one minute on a desktop computer. Fluctuations for the two methods are further discussed in Fig.~\ref{fig:fluctuations}.}
\label{fig:compareSampling}
\end{figure}

\begin{figure}
\includegraphics[width=\linewidth]{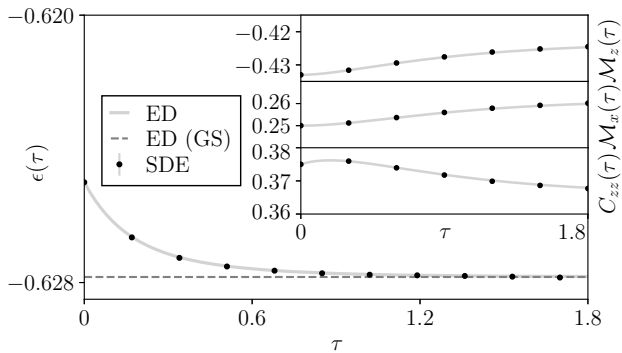}
\caption{Imaginary time evolution for the 2D quantum Ising model. We show the ground state energy (main panel), the longitudinal and transverse magnetization, $\mathcal{M}_z$ and $\mathcal{M}_x$, and the nearest neighbor correlation function $C_{zz}$ (insets, top to bottom) for a $5 \times 5$ system. The system is initialized in the mean field ground state for $\Gamma=1$ and evolved with the same value of $\Gamma$. We compare our results, obtained by solving the SDEs and applying the importance sampling scheme (dots), to exact diagonalization (lines), finding good agreement. At the stopping time $\tau_f =1.8$, the relative error between our estimate of the ground state energy and the true value obtained from ED (horizontal dashed line) is of order $10^{-5}$. Our results were obtained from $5\times 10^7$ realizations with $\Delta \tau=0.01$. The error bars are not visible on the scale of the plot.}
  \label{fig:2Dnumerics} 
 \end{figure}

\begin{figure}
\centering
  \includegraphics[width=\linewidth]{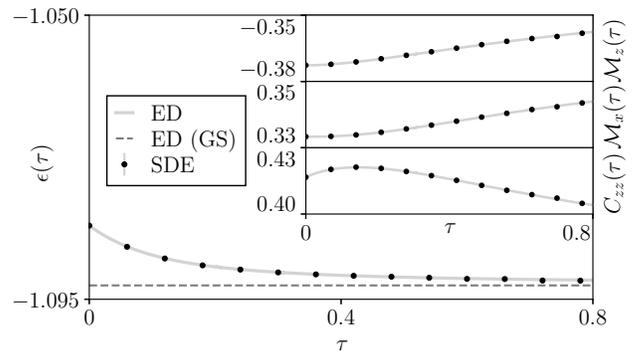}
  \caption{Imaginary time evolution for the 3D quantum Ising model. We consider the same observables of Figs~\ref{fig:1Dnumerics} and \ref{fig:2Dnumerics} for a $3 \times 3 \times 3$ system. The system is initialized in the mean field ground state for $\Gamma=2$ and evolved using the same value of $\Gamma$. Again, the results obtained by solving the SDEs and using importance sampling (dots) are in good agreement with ED (lines). The SDE estimate for the ground state energy at the stopping time $\tau_f=0.8$ is within $0.1 \%$ of the true ground state energy, obtained from ED (dashed horizontal line). Our results were obtained from $7\times 10^7$ realizations of the stochastic process, with $\Delta \tau=0.005$. The error bars are not visible on the scale of the plot.}
\label{fig:3Dnumerics}
\end{figure}

We begin by considering the $D=1$ case, corresponding to the quantum Ising chain. We consider the imaginary time evolution of the ground state longitudinal magnetization $\mathcal{M}_z$, transverse magnetization $\mathcal{M}_x \equiv \sum_{i=1}^N\langle \hat{S}^x_i \rangle /N$ and nearest-neighbor longitudinal correlations $C_{zz} \equiv \sum_{\langle ij \rangle} \langle \hat{S}^z_i \hat{S}^z_j \rangle / N$. As per our general discussion, these quantities are given by Eq.~(\ref{eq:GSexpectationValueStochastic}), where the numerator includes the appropriate stochastic function for each observable. The stochastic functions are given by Eq.~(\ref{eq:longitudinalMagnetization}) for $\mathcal{M}_z$ and by 
\begin{align}
F_{\mathcal{M}_x} &=\frac{F_\mathcal{\mathbbm{1}}}{N} \sum_i \frac{\xi^+_{f,i}(\tau) + \xi^{+*}_{b,i}(\tau)}{1+\xi^+_{f,i}(\tau) \xi^{+*}_{b,i}(\tau)}, \label{eq:transverseMagnetization} \\
F_{C_{zz}} & = \frac{F_\mathcal{\mathbbm{1}}}{N} \sum_{\langle ij \rangle} \left( \frac{1 - \xi^+_{f,i}(\tau) \xi^{+*}_{b,i}(\tau)}{1+\xi^+_{f,i}(\tau) \xi^{+*}_{b,i}(\tau)} \right) \left( \frac{1 - \xi^+_{f,j}(\tau) \xi^{+*}_{b,j}(\tau)}{1+\xi^+_{f,j}(\tau) \xi^{+*}_{b,j}(\tau)}\right) , \label{eq:correlations}
\end{align}
for $\mathcal{M}_x$ and $C_{zz}$ respectively \cite{nonEquilibrium}.
The ground state energy density is obtained from Eqs~(\ref{eq:transverseMagnetization}) and~(\ref{eq:correlations}) as
\begin{align}
\epsilon_G = - \Gamma \mathcal{M}_x - J C_{zz} .
\end{align}
In Fig.~\ref{fig:1Dnumerics} we compare our numerical results for a system of size $N=101$ to imaginary time evolution performed directly in the thermodynamic limit using iTEBD \cite{Vidal2006}. We find excellent agreement across the imaginary time range we consider. The error on the energy estimate $\epsilon(\tau_f)$ we obtain for $\tau_f=4$ is approximately $1.5 \times 10^{-5}$  relative to the exact ground state result. 
To show the improvement of importance sampling according to Eq.~(\ref{eq:sampleAroundSP}) over direct sampling using the naive measure~(\ref{eq:noiseActionDiagonal}), in Fig.~\ref{fig:compareSampling} we compare the performance of the two approaches. 
We fix the physical parameters to $N=15$, $\Gamma=0.4$ and compute the Euclidean time evolution of the energy using the same time step and number of simulations; the results obtained from direct and importance sampling are shown in panels (a) and (b) respectively. It is clear that the importance sampling algorithm produces far better results for the same computational cost. This is further discussed in Section~\ref{sec:fluctuations}, where we study the behavior of fluctuations.
The importance sampling algorithm can be equally applied to higher dimensional systems, for which analytical solutions are typically not available.
The relevant stochastic formulae take a similar form to the one dimensional case \cite{stochasticApproach}. For instance, the stochastic function for the normalization and the longitudinal magnetization are given by Eq.~(\ref{eq:normalization}) and (\ref{eq:longitudinalMagnetization}) respectively, where the sums and products are performed over all lattice sites.
Similarly, the measure transformation is carried out in complete analogy to the one-dimensional case; see Appendix \ref{app:saddlePoint} for further details.
In Fig.~\ref{fig:2Dnumerics}, we consider the 2D quantum Ising model, comparing the results obtained from importance sampling and from exact diagonalization (ED) performed with the QuSpin package~\cite{quSpin}. We consider a $5 \times 5$ system for $\Gamma=1$, performing Euclidean time evolution from the MF ground state. 
Again, we find excellent agreement between our result and ED; the relative error on our estimate for the ground state energy is within $3 \times 10^{-5}$.
Finally, in Fig.~\ref{fig:3Dnumerics} we consider the quantum Ising model in three spatial dimensions. Our results are again in good agreement with ED for a system of size $3\times 3 \times 3$; for the chosen stopping time $\tau_f$, we obtain a relative error on the GS energy within $10^{-3}$. 
We observe that the stopping time $\tau_f$ that can be accessed for a given number of simulations decreases with the dimensionality of the system, due to the faster growth of fluctuations; this is due to the greater connectivity in higher dimensions, and is further investigated in the next Section. 

\subsection{Fluctuations}\label{sec:fluctuations}
\begin{figure}
\centering
 \begin{subfigure}[b]{\linewidth}
         \includegraphics[width=\linewidth]{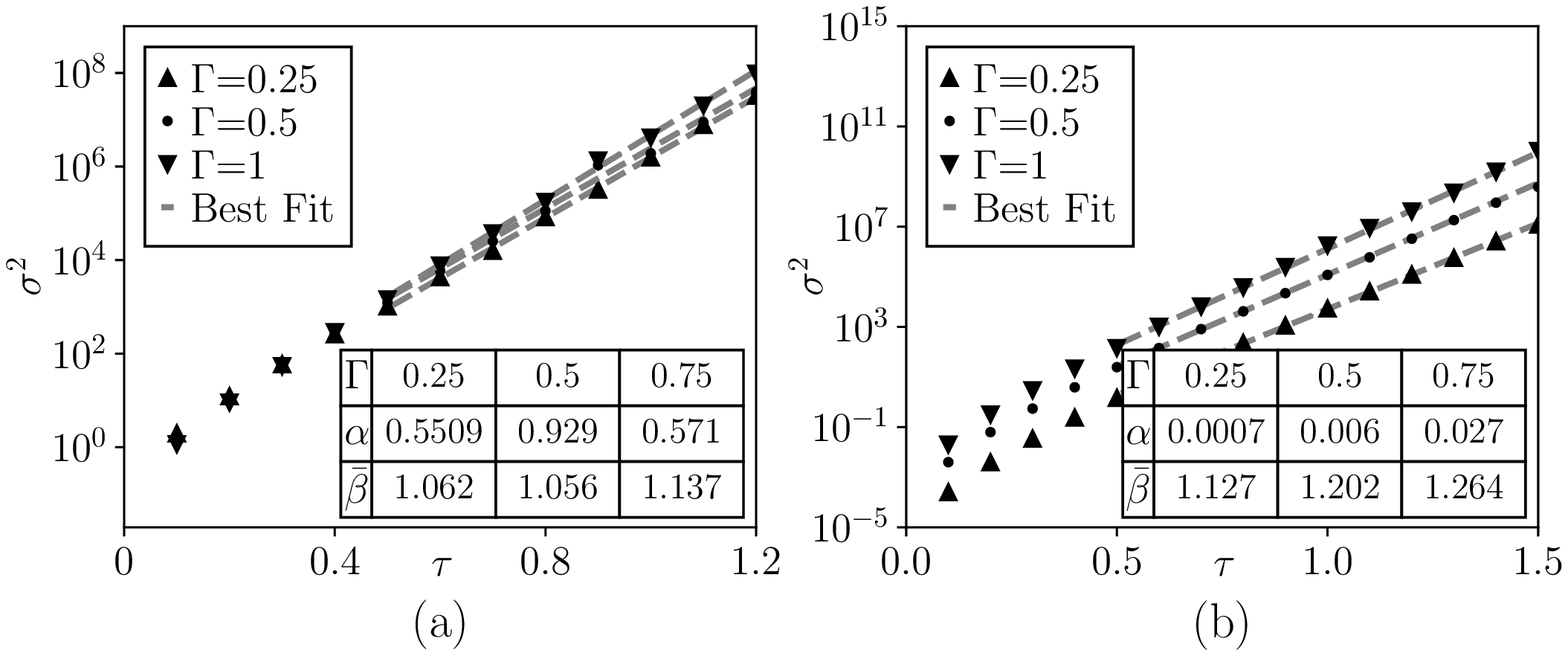}
         \label{fig:fluctNaive}
     \end{subfigure}
 \begin{subfigure}[b]{\linewidth}
          \includegraphics[width=\linewidth]{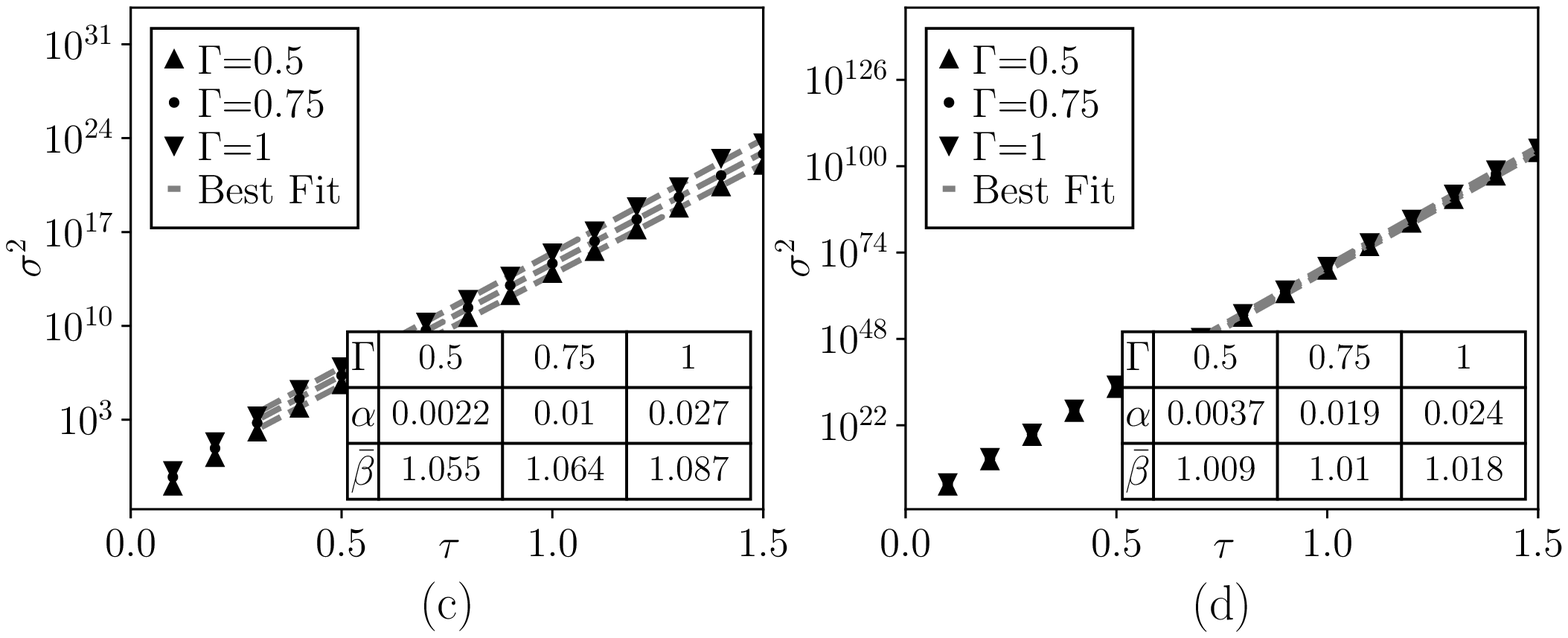}
         \label{fig:fluctNaive}
     \end{subfigure}
  \caption{Growth of fluctuations in the direct and the importance sampling algorithms. Fluctuations are measured by considering the variance $\sigma^2$ of the stochastic function corresponding to the normalization, defined in Eq.~(\ref{eq:normalization}). We begin by considering the one-dimensional quantum Ising chain with $N=7$, sampled using  (a) the direct method and (b) importance sampling. We find that in both cases the behavior of fluctuations is well approximated by Eq.~(\ref{eq:fitFunction}) with $\bar{\beta} \equiv \beta/2ND \approx 1$. However, the prefactor $\alpha$ multiplying the exponential is orders of magnitude smaller for importance sampling than for direct sampling. In panels (c-d) we extend this analysis to (c) a $3\times3$ and (d) a $3 \times 3\times 3$ quantum Ising model. We consider only the importance sampling method, as the rapid growth of fluctuations would make it difficult to gather sufficient statistics for direct sampling. We find that the functional form~(\ref{eq:fitFunction}) accurately describes the growth of fluctuations also in higher dimensions. For importance sampling, in all cases (b-d) the prefactor $\alpha$ gradually increases from zero as $\Gamma$ is increased. Data about the direct approach, panel (a), were obtained from $3 \times 10^7$ simulations. Data about the importance sampling approach, panels (b-d), were obtained from $10^5$ simulations; these were sufficient, due to the smaller extent of fluctuations. }
\label{fig:fluctuations}
\end{figure}

Having demonstrated the applicability of the method to higher dimensional systems, we now turn to investigating its numerical performance, quantitatively comparing the direct and importance sampling schemes. 
Fluctuations in the stochastic quantities play a central role: for a given number of simulations, the growth of fluctuations ultimately determines the time scale beyond which physical results are not correctly reproduced. Therefore, an increasing number of simulations is needed as the stopping time is increased.
The central limit theorem implies that the fluctuations in the observable $\mathcal{O}$ computed from the stochastic approach are determined by the variance $\sigma^2$ of the corresponding stochastic quantity $f_\mathcal{O}$~\cite{nonEquilibrium}:
\begin{align}
\sigma^2(f_\mathcal{O}) \equiv \langle |f_\mathcal{O}|^2\rangle_\phi - |\langle f_\mathcal{O} \rangle_\phi|^2 .
\end{align}
The variance $\sigma^2$ is therefore directly related to the number of simulations required to obtain a given accuracy.
We illustrate this by considering the normalization function~(\ref{eq:normalization}): the behavior of this quantity is found to be representative of other observables, due to the similar functional form of the corresponding stochastic functions; see for example Eq.~(\ref{eq:longitudinalMagnetization}).
In the classical case of the $D$-dimensional Ising model with $\Gamma=0$, the SDEs~(\ref{eq:ITising}) are exactly solvable. For direct sampling, one obtains
\begin{align}\label{eq:varianceClassical}
\sigma^2(\tau)= e^{2NDJ\tau} - e^{NDJ\tau}
\end{align}
where $ND$ is the total number of interactions in the system.
In contrast, the variance $\sigma^2$ for the importance sampling scheme vanishes identically and a single trajectory is sufficient to give the exact result. For finite $\Gamma$, the behavior of fluctuations in the two approaches can be investigated numerically. As shown in Fig.~\ref{fig:fluctuations}, we find that this is captured by the functional form
\begin{align}\label{eq:fitFunction}
\sigma^2 = \alpha e^{\beta \tau},
\end{align} 
with $\beta \approx 2 D N $. Thus, the exponential growth of fluctuations with $N$, $D$ and $\tau$, which we found for direct sampling in the classical case $\Gamma=0$, survives for finite $\Gamma$, and also applies to importance sampling. 
This is consistent with the numerical analysis carried out in Ref.~\cite{nonEquilibrium} for real time evolution, and with the argument of Refs~\cite{ringelThesis,ringelGritsev} suggesting that a large deviation principle may be at play with respect to the system size $N$. However, direct and importance sampling differ substantially in the prefactor $\alpha$ multiplying the exponential. We find that $\alpha$ depends heavily on $\Gamma$, as shown in the tables of Fig.~\ref{fig:fluctuations}. 
For direct sampling, one has $\alpha = O(1)$ for all $\Gamma$. On the other hand, for importance sampling, $\alpha$ gradually increases from zero as $\Gamma$ is increased. For small to intermediate field strengths $\Gamma \approx O(1)$, $\alpha$ can be orders of magnitude smaller for importance sampling than for direct sampling. 
Thus, the importance sampling scheme can strongly suppress the growth of fluctuations with time and the system size. This significantly extends the regime of applicability of the stochastic method before fluctuations become sizable, although it does not eliminate their exponential growth.
It would be interesting to clarify the relation between stochastic fluctuations and entanglement. The importance sampling method completely eliminates fluctuations when the true ground state is a product state and does not have entanglement, as in the classical limit. Thus, both the presence of residual fluctuations and the growth of entanglement signal the departure from a product state; whether a direct connection between these exists will be investigated in future work.

The computational cost of the numerical stochastic approach is mainly determined by the growth of fluctuations. 
The runtime of a given stochastic simulation scales linearly with $N$, $\tau_f$ and inversely with the time step $\Delta \tau_f$. Such simulations are straightforwardly parallelized, as they feature independent trajectories. For instance, the results of Fig.~\ref{fig:1Dnumerics} were computed from $10^3$ batches of $10^5$ independent simulations each; each batch takes approximately $0.8$ hours on $16$ cores. 
For fixed computational resources, the observed growth of fluctuations with $\tau$ and $N$ then leads to a trade-off between accessible time scales and system sizes.
Thus, further developments will be needed for the numerical approach to be effectively applicable to late times and large systems.
However, the significant suppression of fluctuations achieved using importance sampling could be particularly interesting in view of real time applications: a generalization of this method might prove useful in settings where existing techniques are significantly limited, such as higher dimensions, to explore intermediate time and system size regimes before fluctuations become dominant. The real time numerical approach~\cite{stochasticApproach,nonEquilibrium} is indeed fully analogous to the present imaginary time case, suggesting that the approach may readily generalize; however, a key difference is that for real time evolution it is not possible to make use of the $\tau \rightarrow \infty$ limit to simplify the saddle point equation, and a different method (e.g. recursion) will generally be needed in order to obtain the saddle point configuration.

In summary, in this Section we used the disentanglement approach to numerically compute ground state expectation values as averages over classical stochastic trajectories.
This approach is formally exact, and can be made more efficient by employing an importance sampling scheme, based on preferentially sampling trajectories close to the relevant saddle point configuration. 
Notably, this technique can be applied regardless of dimensionality.
The proposed stochastic approach provides an alternative numerical method for the evaluation of ground state expectation values, which is conceptually different from existing techniques. For instance, in worldline quantum Monte Carlo (MC) methods~\cite{Suzuki1977,Prokofev1996,Prokofev1998,Troyer2009}, one splits the Hamiltonian into a diagonal, ``classical'' term, and a non-diagonal term; the latter is expanded in time-dependent perturbation theory, so that one is left with a sum of classical integrals which can be evaluated by MC. In contrast, here the sampling is performed over deviations from the mean field trajectory. 
Other classes of MC methods for spins include the previously mentioned diagrammatic approaches~\cite{KulaginPRB2013,KulaginPRL2013,huang2016}, based on mapping to fermions, or stochastic series expansions, whereby the whole Hamiltonian is treated perturbatively~\cite{Sandvik1991,Troyer2009}.
The numerical performance of the stochastic approach is currently inferior to well-established numerical techniques for ground states, such as quantum MC~\cite{Creswick1988,Blote2002} or tensor network approaches~\cite{Jordan2008}. 
However, the importance sampling method substantially extends the regime of applicability of the stochastic approach by mitigating fluctuations. This suggests that the rate of growth of fluctuations is not intrinsically fixed, and that it could be possible to suppress it even further by means of appropriate sampling schemes or approximations.

\section{Conclusions}\label{sec:conclusion}
In this manuscript, we showed that the disentanglement formalism \cite{hoganChalker,galitski,ringelGritsev,stochasticApproach,nonEquilibrium} provides a broadly applicable framework to describe many-body quantum spin ground states, bridging concepts from lattice spin systems, field theory and classical stochastic processes.
In this approach, expectation values are exactly expressed as functional integrals over scalar fields, amenable to both analytical treatment and numerical evaluation.

Considering the quantum Ising model in $D$ spatial dimensions, we showed that the leading mean-field contribution to observables corresponds to the saddle point of a suitable effective action. Analytical corrections beyond mean field are then computed by expanding the action about the saddle point to a desired order. Within this approach, quantum phase transitions are associated to the expansions about different saddle points abruptly swapping their role in providing the dominating contribution to the ground state energy. It would be interesting to investigate how the proposed picture generalizes in the presence of more complicated phase diagrams.

In addition, we showed that the disentanglement method can alternatively be used as a numerical tool to compute ground state expectation values from classical stochastic processes. 
The main drawback of the numerical approach is the exponential growth of fluctuations in the stochastic quantities with time and the system size, as previously found for real-time applications~\cite{stochasticApproach,nonEquilibrium}.
Analogous exponential bottlenecks are often encountered in quantum many-body physics, and can sometimes be circumvented. Examples are the Monte Carlo sign problem~\cite{Henelius2000,Troyer2005,marvian2019}, which in certain systems is eliminated through basis changes~\cite{Nakamura1998,Alet2016,hann2017,Wessel2017}, or the exact contraction of 2D tensor networks~\cite{schuch2007}, which can be replaced by more efficient approximate schemes~\cite{Jiang2008,pizorn2011,wang2011}.
Similarly, it might be possible to suppress the growth of fluctuations in the stochastic approach by means of suitable approximations or sampling schemes. As a promising step in this direction, we introduced an importance sampling numerical technique, capable of significantly mitigating the growth of fluctuations by biasing the measure towards the saddle point trajectory. 
An interesting direction for future research would then be investigating whether basis changes or approximate approaches can eliminate or further suppress the exponential growth of fluctuations.
Notably, the present method also applies in higher dimensions, as we showed by considering the 2D and 3D quantum Ising model. A real-time generalization of this approach might then prove useful to study the non-equilibrium dynamics of higher-dimensional quantum systems, where existing techniques are far less effective than for ground states. 
In this context, the disentanglement method can provide an analytical formulation from which to develop approximations, as well as a numerical tool: although fluctuations are likely to eventually prevail, a suitable importance sampling scheme might still be able to access regimes beyond the reach of currently available techniques.

Several directions for further development of the method can be envisaged, including cluster approaches~\cite{Ferreira1977,Oliveira1991,jin2013,Zimmer2016} or establishing connections to tensor networks~\cite{Schollwock2011,Orus2014}. The direct relation between exact equations and numerical sampling afforded by the disentanglement formalism might also aid the development of problem-specific approximations, based on the physical understanding of a given system. Approximate analytical and numerical approaches could prove useful in studying systems that pose severe challenges to existing techniques, such as frustrated magnets~\cite{hoganChalker,lacroix2011introduction}.

{\em Acknowledgments.}--- SDN would like to thank M. J. Bhaseen, J. Chalker, B. Doyon, V. Gritsev, A. Lamacraft, A. Michailidis and M. Serbyn for helpful feedback and stimulating conversations.
SDN acknowledges funding from the Institute of Science and Technology (IST) Austria, and from the European Union's Horizon 2020 research and innovation programme under the Marie Sk\l{}odowska-Curie Grant Agreement No. 754411. SDN also acknowledges funding from the EPSRC Centre for Doctoral Training in Cross-Disciplinary Approaches to Non-Equilibrium Systems (CANES) under grant EP/L015854/1. 

%

\appendix 
\section{Disentanglement Transformation}\label{app:stochasticDecoupling}
In order to make the manuscript self-contained, in this Appendix we recapitulate the key steps of the disentanglement formalism, focusing on imaginary time dynamics and providing additional details on the higher-dimensional case.
\subsection{General Case}\label{app:dis1D}
In the disentanglement approach, the Euclidean time evolution operator corresponding to the Hamiltonian (\ref{eq:hamiltonian}),
\begin{align}\label{eq:Uapp}
\hat{U}(\tau) \equiv e^{-\tau \hat{H}}  = e^{ \tau ( J \sum_{ijab} \mathcal{J}^{ab}_{ij} \hat{S}^a_i \hat{S}^b_j + \sum_{ia} h^a_i \hat{S}^a_i ) },
\end{align}
is expressed as
\begin{equation}\label{eq:evOptrApp}
\hat{U}(\tau) =  \int \mathcal{D} \varphi
\mathrm{e}^{-S_0[\varphi]}  \prod_i  e^{\xi_i^+(\tau)\hat
  S_i^+}e^{\xi_i^z(\tau)\hat S_i^z}e^{\xi_i^-(\tau)\hat
  S_i^-} ,
\end{equation}
where $S_0$ is given by Eq.~(\ref{eq:noiseAction}) and the disentangling variables $\xi^a_i$ satisfy Eqs~(\ref{eq:SDEs}). Eq.~(\ref{eq:evOptrApp}) is obtained in a two-step process \cite{hoganChalker,galitski,ringelGritsev,stochasticApproach}. First, interactions are decoupled thanks to the Hubbard-Stratonovich transformation \cite{stratonovich,hubbard}. Following a Trotter decomposition of the exponential in Eq.~(\ref{eq:Uapp}), at each time slice one has
\begin{align}\label{eq:optrHS}
\begin{split}
& e^{\Delta \tau J \sum_{ijab} \mathcal{J}^{ab}_{ij} \hat{S}^a_i \hat{S}^b_j } = \\
& \mathcal{C}  \int \prod_{ai} \mathrm{d}\varphi_{i}^a e^{- \frac{1}{4J } \Delta \tau \sum_{ijab}(\mathcal{J}^{-1})^{ab}_{ij}\varphi^a_i \varphi^b_j + \Delta \tau \sum_{aj} \varphi^a_j \hat{S}^a_j } 
\end{split},
\end{align}
where $\mathcal{C}$ is a normalization constant and we neglected terms $O(\Delta\tau^2)$. Eq.~(\ref{eq:optrHS}) is an operatorial identity; the fields $\varphi^a_i$ are in general complex and their integration range in the complex plane is chosen in such a way as to make the integral in Eq.~(\ref{eq:optrHS}) convergent~\cite{nonEquilibrium}.
It is convenient to rescale the fields as $\varphi^a_i \rightarrow J \varphi^a_i$, in order to make them dimensionless. 
Applying this rescaling and taking the continuum limit, Eq.~(\ref{eq:optrHS}) yields
 \begin{equation}\label{eq:evOptrstep1}
\hat{U}(\tau) =  \int \mathcal{D} \varphi
\mathrm{e}^{-S_0[\varphi]} \mathbb{T} \prod_i e^{\sum_{a} \int_0^\tau [h^a_i(\tau^\prime) + J \varphi^a_i(\tau^\prime)] \hat{S}^a_i \dd \tau^\prime} ,
\end{equation}
where the symbol $\mathbb{T}$ denotes time ordering.
Eq.~(\ref{eq:evOptrstep1}) describes a system of non-interacting spins under the effect of complex valued stochastic fields $\varphi^a_i$ \cite{hoganChalker,ringelGritsev}. Since interactions are decoupled inside the integral, the evolution of each spin occurs over its (complexified) Bloch sphere. Time-ordered exponentials can then be expressed in terms of ordinary exponentials by means of a Lie-algebraic disentanglement transformation, also known as Wei-Norman-Kolokolov transformation \cite{weiNorman,kolokolov,hoganChalker,galitski,ringelGritsev}. Namely, at each lattice site one has
\begin{align}\label{eq:disentanglement}
 \mathbb{T}  e^{\sum_{a} \int_0^\tau [h^a_i(\tau^\prime) + J \varphi^a_i(\tau^\prime)] \hat{S}^a_i \dd \tau^\prime} = e^{\xi_i^+(\tau)\hat   S_i^+}e^{\xi_i^z(\tau)\hat S_i^z}e^{\xi_i^-(\tau)\hat  S_i^-} .
\end{align}
This amounts to parameterizing the trajectory of each spin on its Bloch sphere in terms of a set of coordinates $\xi^a_i$, termed the disentangling variables. Eq.~(\ref{eq:disentanglement}) can be seen as the defining equation of $\xi^a_i$; differentiating both sides of~(\ref{eq:disentanglement}) and equating the coefficients that multiply the spin operators yields the SDEs~(\ref{eq:SDEs}). Alternatively, these can be obtained from differential geometry \cite{ringelGritsev}. The initial conditions $\xi^a_i(0)=0$ are fixed by the requirement $\hat{U}(0) = \mathbbm{1}$. The discussion of this Section can be readily generalized to the modified time evolution operator $\hat{\mathcal{U}}(\tau)$, given by Eq.~(\ref{eq:modifiedU}); the initial conditions of the disentangling variables are then given by~(\ref{eq:initialConditions}).

\subsection{Details on the Disentanglement Transformation in Higher Dimensions}
While the formalism outlined in the previous Section is fully general, in this Section we show in greater detail how the disentanglement transformation works in higher dimensional settings of particular physical interest, providing useful formulae for analytical and numerical applications. 
Let us consider a Hamiltonian describing a system on a $D$-dimensional hypercubic lattice:
\begin{align}\label{eq:hamiltonianD}
\hat{H}= - \sum_{\boldsymbol{i} \boldsymbol{j}} \sum_{ab} \mathcal{J}^{ab}_{\boldsymbol{i} \boldsymbol{j}} \hat{S}^a_{\boldsymbol{i}} \hat{S}^b_{\boldsymbol{j}}   - \sum_{\boldsymbol{i}} \sum_a h^a_{\boldsymbol{i}} \hat{S}^a_{\boldsymbol{i}} .
\end{align}
We focus on the case where $\mathcal{J}$ only couples spins along the lattice axes, i.e. the sites $\boldsymbol{i}$, $\boldsymbol{j}$ coupled by $\mathcal{J}_{\boldsymbol{i} \boldsymbol{j}}$ differ by a single index, $i_d \neq j_d$. For this choice of $\mathcal{J}$, one has
\begin{align}
\mathcal{J}^{ab}_{\boldsymbol{i} \boldsymbol{j}} = \sum_d J_d (\mathcal{J}^d)^{ab}_{i_d j_d} \prod_{d^\prime\neq d} \delta_{i_{d^\prime} j_{d^\prime}} ,
\end{align}
where $J_d$ are interaction strengths and the matrices $\mathcal{J}^d$ couple spins along the dimension $d$. $\mathcal{J}^d$ can be seen as $3N_d \times 3N_d$ matrices $\mathcal{J}^d_{\alpha\beta}$ by introducing multi-component indices $\alpha=\{i_d,a\}, \beta=\{j_d,b \}$. One has
\begin{align}
\sum_{\boldsymbol{i} \boldsymbol{j}} \sum_{ab} \mathcal{J}_{\boldsymbol{i} \boldsymbol{j}}^{ab} \hat{S}^a_{\boldsymbol{i}} \hat{S}^b_{\boldsymbol{j}} = \sum_{d=1}^D J_d \sum_{\boldsymbol{i}} \hat{S}^a_{\boldsymbol{i}} \sum_{j_d} (\mathcal{J}^d)^{ab}_{i_d j_d}  \hat{S}^b_{i_1 \dots j_d \dots i_D}  .
\end{align} 
Exponentiating the interaction term in~(\ref{eq:hamiltonianD}) and considering an infinitesimal time slice, one obtains
\begin{align}\label{eq:optrHSD}
\begin{split}
&e^{ \Delta \tau \sum_{\boldsymbol{i} \boldsymbol{j}ab} \mathcal{J}^{ab}_{\boldsymbol{i} \boldsymbol{j}} \hat{S}^a_{\boldsymbol{i}} \hat{S}^b_{\boldsymbol{j} }} \\
& =  \prod_{d=1}^D\prod_{\boldsymbol i \neq i_d} e^{\Delta \tau  \sum_{i_d j_d ab} J_d(\mathcal{J}^d)^{ab}_{i_d j_d} \hat{S}^a_{\boldsymbol{i}} \hat{S}^b_{i_1 \dots j_d \dots i_D}} .
\end{split}
\end{align}
We can apply the HS transformation to each term:
\begin{widetext}
\begin{align}
e^{ \Delta \tau  \sum_{i_d j_d a b } J_d(\mathcal{J}^d)^{ab}_{i_d j_d} \hat{S}^a_{\boldsymbol{i}} \hat{S}^b_{i_1 \dots j_d \dots i_D} } = \mathcal{C} \int \prod_{i_d=1}^{N_d} \left( \mathrm{d}\varphi^d_{\boldsymbol{i}} \right) & e^{-\frac{\Delta \tau}{4 J_d	} \sum_{i_d j_d a b }  [(\mathcal{J}^d)^{-1}]^{ab}_{i_d j_d} (\varphi^d)^a_{\boldsymbol{i}} (\varphi^d)^b_{i_1 \dots j_d \dots i_D} + \Delta \tau \sum_{j_d a} (\varphi^d)^a_{i_1 \dots j_d \dots i_D} \hat{S}^a_{i_1\dots j_d \dots i_D}}.
\end{align}
\end{widetext}
Taking the continuum limit and rescaling $\varphi^d \rightarrow J_d \varphi^d $, this yields
\begin{align}\label{eq:decoupledD}
e^{-\tau \hat{H}} = \int\mathcal{D}\varphi  e^{-S_0 [\varphi]}  \mathbb{T} e^{ \int_0^\tau \mathrm{d}\tau \left[ \sum_{d} J_d \sum_{\boldsymbol{i}a} (\varphi^d)^a_{\boldsymbol{i}} +h^a_{\boldsymbol{i}} \right]\hat{S}^a_{\boldsymbol{i}}  } 
\end{align}
where the noise action in $D$ dimensions is given by
\begin{align}\label{eq:noiseActionD}
S_0[\varphi]= \frac{1}{4} \int_0^\tau \mathrm{d}\tau \sum_d J_d \sum_{\boldsymbol{i}j_d ab}[(\mathcal{J}^{d})^{-1}]^{ab}_{i_d j_d}(\varphi^d)^{a}_{\boldsymbol{i}} (\varphi^d)^{b}_{i_1 \dots j_d i_D} .
\end{align}
It can be seen that for a $D$ dimensional system with $N= N_1 \times \dots \times N_d$ spins, one needs in general to introduce $3DN$ Hubbard-Stratonovich fields.
The individual time ordered exponentials in Eq.~(\ref{eq:decoupledD}) can then be expressed in terms of ordinary exponentials, as done in Appendix~\ref{app:dis1D}.
As in the 1D case, a change of variables can be performed to make the noise action $S_0$ diagonal. Let us introduce the notation $\bar{\boldsymbol{i}}_d \equiv i_1 \dots i_{d-1} i_{d+1} \dots i_D$ and $(\varphi^d)^a_{\boldsymbol{i}} = (\varphi^d)_{\bar{\boldsymbol{i}}_d \alpha}$ with $\alpha=\{i_d,a \}$. Eq.~(\ref{eq:noiseActionD}) is then diagonalized by the transformation
\begin{align}\label{eq:diagonalizeNoiseD}
(\varphi^d)^a_{\boldsymbol{i}}= (\varphi^d)_{\bar{\boldsymbol{i}}_d \alpha} = \sum_{\beta=1}^{3 N_d} (O^d)_{\alpha \beta}(\phi^d)_{\bar{\boldsymbol{i}}_d \beta}  ,
\end{align}
where $O^d$ is a $3N_d \times 3N_d$ matrix defined as for the 1D case, but in terms of $\mathcal{J}^d$. Using Eq.~(\ref{eq:diagonalizeNoiseD}), we obtain
\begin{align}
\begin{split}
& e^{\tau \sum_{\boldsymbol{i} \boldsymbol{j}} \mathcal{J}_{\boldsymbol{i} \boldsymbol{j}} \hat{S}^z_{\boldsymbol{i}} \hat{S}^z_{\boldsymbol{j} }}  \\
& = 
\int \mathcal{D}\phi e^{-S_0[\phi] + \int_0^\tau \mathrm{d}\tau \sum_{\boldsymbol{i}a}   \hat{S}^a_{\boldsymbol{i}} \sum_{d} J_d \sum_{\beta=1}^{3 N_d} (O^d)_{\alpha \beta}(\phi^d)_{\bar{\boldsymbol{i}}_d \beta}} 
\end{split}
\end{align}
with 
\begin{align}
S_0[\phi]= \frac{1}{2} \int_0^{\tau_f} \mathrm{d}\tau \sum_{d}\sum_{\boldsymbol{i}a} [(\phi^d)^a_{\boldsymbol{i}}]^2.
\end{align}

\section{Euclidean Time Dynamics}\label{app:isingSDEs}
In this Appendix we study the Euclidean time dynamics of the disentangling variables~(\ref{eq:SDEs}), which fully encode the quantum system.
\subsection{Ising SDEs}
The stochastic representation (\ref{eq:stochasticTimeEvol}) of the Euclidean time evolution operator is formally exact; this implies that the statistics of the classical disentangling variables $\xi=\{\xi^a_i\}$ contain all the information about the corresponding quantum problem. In the case of real time evolution, this observation was drawn upon in Refs \cite{stochasticApproach,nonEquilibrium} to numerically investigate the relation between fluctuations in the disentangling variables and dynamical quantum phase transitions \cite{heyl2013,heyl2018}.
Here we consider the imaginary time behavior of the disentangling variables, which encodes all information about the ground state of the corresponding quantum problem.
For definiteness, we consider the quantum Ising model, given by the Hamiltonian~(\ref{eq:Ising}). For the one-dimensional quantum Ising chain, the general result (\ref{eq:SDEs}) specializes to the Euclidean Ising SDEs \cite{ringelGritsev,stochasticApproach,nonEquilibrium}
\begin{subequations}\label{eq:ITisingApp}
\begin{align}
\dot{\xi}^+_i(\tau)&= \frac{\Gamma}{2} (1-{\xi^+_i}^2) + J \xi^+_i  \sum_j O_{ij} \phi_j  \label{eq:ITising1App} , \\
\dot{\xi}^z_i(\tau) &= -\Gamma \xi^+_i  + J \sum_j O_{ij} \phi_j , \label{eq:ITising2App}  \\
\dot{\xi}^-_i(\tau) &= \frac{\Gamma}{2}\exp{\xi^z_i}, \label{eq:ITising3App} 
\end{align}
\end{subequations}
which are here expressed in terms of the fields $\phi_i$ that diagonalize the noise action~(\ref{eq:noiseAction}).

\subsection{Exactly Solvable Limits}
To the best of our current knowledge, Eqs~(\ref{eq:ITisingApp}) are only exactly solvable in the classical ($\Gamma=0$) and non-interacting ($J=0$) cases. This was discussed in Ref.~\cite{nonEquilibrium} for real time evolution and in the special case $\xi^a_i(0)=0$; here we consider Euclidean time and general initial conditions, as it is relevant for our current purposes. For the present discussion, we focus on the one-dimensional case, which is sufficient to illustrate the relevant properties of the disentangling variables; the higher-dimensional version of Eq.~(\ref{eq:ITisingApp}) is given by Eq.~(\ref{eq:highDising}) in Appendix~\ref{app:saddlePoint}.
In the classical case with $\Gamma=0$, the non-linear term in Eq.~(\ref{eq:ITising1App}) vanishes and $\xi^+_i$ performs driftless geometric Brownian motion. This is exactly solvable, giving
\begin{align}
\xi^+_i(\tau) = \xi^+_i(0) \exp\left[ \sum_j O_{ij}\int_0^\tau \phi_j(s)\mathrm{d}s\right] ,
\end{align}
where we used $(OO^T)_{ii}\propto \mathcal{J}_{ii}=0$. In the classical limit, $\xi^z_i$ is decoupled from $\xi^+_i$ and satisfies Brownian motion:
\begin{align}
\xi^z_i(\tau) = \xi^z_i(0)+ \int_0^\tau \sum_j O_{ij} \phi_j(s) \dd s ,
\end{align}
while $\xi^-_i(\tau)=\xi^-_i(0)$. 

In the non-interacting limit $J=0$, Eqs~(\ref{eq:ITisingApp}) become deterministic and solvable, yielding
\begin{subequations}
\begin{align}
\xi^+_i(\tau) &= \xi^+_i(0) + \frac{1-\xi^{+2}_i(0)}{\xi^+_i(0)+\coth ( \Gamma \tau /2)} \label{eq:xpNI} , \\
\xi^z_i(\tau) &= \xi^z_i(0) -2 \log \left[ \cosh(\Gamma \tau/2) +\xi^+_i(0) \sinh(\Gamma \tau/2)\right] ,\\
\xi^-_i(\tau) &=  \xi^-_i(0) +\frac{\exp[\xi^z_i(0)] }{\xi^+_i(0) + \coth(\Gamma\tau/2)} .
\end{align}
\end{subequations}
\vspace{1cm}
\subsection{Moments of the Disentangling Variables}\label{app:moments}
In the general case with finite $\Gamma$ and $J$, the SDEs~(\ref{eq:ITisingApp}) cannot be solved exactly to the best of our knowledge. However, analytical insights about~(\ref{eq:ITisingApp}) can still be obtained. Of particular interest is the behavior of the variables $\xi^+_i$: as observed in Refs~\cite{ringelGritsev,stochasticApproach,nonEquilibrium}, these play a key role, being the primary source of non-linearity in~(\ref{eq:ITisingApp}) (the variable $\xi^-_i$ is seldom needed to compute observables) and the sole disentangling variable whose equation of motion is autonomous, not involving any other $\xi^a_i$.
The stationary probability distribution attained at late times by $\xi^+_i(\tau)$ was obtained in Refs~\cite{ringelThesis,ringelGritsev}. Additional information is encoded in the moment-generating function $G_i(\lambda,\tau)$ of $\xi^+_i(\tau)$, which satisfies $\partial^n_\lambda G_i(\lambda,\tau)\vert_{\lambda=0} = \langle \xi^{+n}_i(\tau) \rangle_\phi$ and gives access to the Euclidean time-dependent moments of $\xi^+_i$.
To compute this, we define the stochastic function $g_i(\lambda,\tau) \equiv  e^{\lambda \xi^+_i(\tau)}$, such that $G_i(\lambda,\tau) \equiv \langle g_i(\lambda,\tau)  \rangle_\phi$. The equation of motion of $g_i(\tau)$ is obtained by applying the Ito chain rule \cite{ito,kloeden}:
\begin{widetext}
\begin{equation}
\frac{\mathrm{d}}{\mathrm{d}\tau} g_i(\lambda,\tau) = \lambda \frac{\Gamma}{2} \left(1-{\xi^+_i}^2\right) g_i(\lambda,\tau) + \lambda \xi^+_i \sum_j O_{ij} \phi_j  g_i(\lambda,\tau)+ \frac{1}{2}\lambda^2 {\xi^+_i}^2 \sum_j O_{ij} O_{ij} g_i(\lambda,\tau) .
\end{equation}
\end{widetext}
It can be easily shown that the matrix $O O^T$ is proportional to $\mathcal{J}$ and hence has no diagonal term \cite{nonEquilibrium}; this implies that the Ito drift term proportional to $ \sum_j O_{ij} O_{ij}$ gives no contribution. It is also convenient to write $\xi^+_i g_i = \frac{\partial}{\partial \lambda } g_i$. With these simplifications, we obtain 
\begin{equation}\label{eq:eomGenFn}
\frac{\mathrm{d}}{\mathrm{d}\tau} g_i(\lambda,\tau) = \Big[\lambda \frac{\Gamma}{2} \left(1-\frac{\partial^2}{\partial \lambda^2}\right) +  \sum_j O_{ij} \phi_j \frac{\partial}{\partial \lambda}    \Big] g_i(\lambda,\tau) .
\end{equation}
Considering the expectation value of Eq.~(\ref{eq:eomGenFn}) and using the property of Ito calculus $\langle g_i(\tau) \phi_j(\tau)\rangle_\phi =0$ $\forall\ i, j$ we obtain the partial differential equation satisfied by the moment-generating function:
\begin{equation}\label{eq:eomG}
\frac{\partial}{\partial t} G_i(\lambda,\tau) = \Big[\lambda \frac{\Gamma}{2} \left(1-\frac{\partial^2}{\partial \lambda^2}\right) \Big] G_i(\lambda,\tau) ,
\end{equation}
with initial conditions $G_i(0,\tau)=G_i(\lambda,0)=1$. 
Eq.~(\ref{eq:eomG}) can be solved exactly, yielding
\begin{equation}\label{eq:genFunctionImag}
G_i(\lambda,\tau) =  \exp \left[ \lambda \left(\xi^+_i(0) + \frac{1-\xi^{+2}_i(0)}{\xi^+_i(0)+\coth ( \Gamma \tau /2)}\right) \right].
\end{equation}
This result predicts that all moments of $\xi^+_i(\tau)$ are given by powers of the deterministic trajectory obtained in the non-interacting case with $J=0$: the moments of each individual $\xi^+_i$ contain no information about the interacting quantum system. All information is therefore encoded in the correlations between variables at different sites. 
We note that the findings of the present Section do not apply to real time evolution: in that case, the moments of $\xi^+_i(t)$ for non-zero $J$ differ from the non-interacting result. This discrepancy can be traced back to the failure of the analytic continuation of Eq.~(\ref{eq:genFunctionImag}) to real time.
\subsection{Joint Probability Distribution}
Since the information about interactions is contained in the joint statistics of the $\xi^+_i$ variables, we investigate the joint probability distribution $P[\xi^+]\equiv P[\{\xi^+_i\}]$. The stochastic process $\xi^+_i$ has drift and diffusion 
\begin{subequations}\label{eq:driftAndDiffusion}
\begin{align}
a_i(\xi^+_i) &= \frac{\Gamma}{2} (1-\xi^{+2}_i)\label{eq:drift} , \\
B_{ij}(\xi^+_i) &= \xi^+_i O_{ij} \label{eq:diffusion} 
\end{align}
\end{subequations}
respectively. The probability distribution of its realizations is given by \cite{tirapegui,arnold1,arnold2,ringelGritsev}
\begin{align}\label{eq:jointProbability}
P[\xi^+] = \mathcal{C}_\xi e^{- I[\xi^+]} ,
\end{align}
where $\mathcal{C}_\xi$ is a normalization constant and
\begin{align}
I[\xi^+] &= \int_0^\tau \mathrm{d}\tau^\prime L(\xi^+,\dot{\xi}^+) \label{eq:jointProbabilityAction} , \\
L(\xi^+,\dot{\xi}^+) &= \frac{1}{2} \sum_{ij} [\dot{\xi}^+_i-a_i(\xi^+_i)]  \mathcal{B}^{-1}_{ij}(\xi^+)  [\dot{\xi}_j-a_j(\xi^+_j)], \\
\mathcal{B}_{ij}(\xi^+) &= \sum_k B_{ik}(\xi^+_i) B_{jk}(\xi^+_j) .
\end{align}
Eqs~(\ref{eq:driftAndDiffusion}) give $\mathcal{B}_{ij}(\xi^+)= 2J \mathcal{J}_{ij} \xi^+_i \xi^+_j $ and 
\begin{align}\label{eq:actionXp}
\begin{split}
& L(\xi^+,\dot{\xi}^+) = \\
& \frac{1}{4J} \sum_{ij} \frac{1}{\xi^+_i \xi^+_j} \left[ \dot{\xi}^+_i-\frac{\Gamma}{2}(1-{\xi^+_i}^2) \right]  \mathcal{J}^{-1}_{ij} \left[\dot{\xi}^+_j-\frac{\Gamma}{2}(1-{\xi^+_j}^2)\right].
\end{split}
\end{align}
Eq.~(\ref{eq:jointProbability}) provides the measure when the stochastic expression for an observable is expressed as a path integral over the variables $\xi^+_i$ rather than the fields $\phi_i$ \cite{ringelGritsev}.
When sampling according to the distribution (\ref{eq:jointProbability}), the likeliest trajectory is obtained by extremizing the weight $I[\xi^+]$ with respect to $\xi^+_i(\tau)$. By solving the corresponding Euler-Lagrange equations, we readily see that the dominant trajectory is the non-interacting solution~(\ref{eq:xpNI}). Therefore, when applying the stochastic approach using direct sampling~\cite{stochasticApproach,nonEquilibrium}, one typically samples trajectories which are nearly non-interacting. 

We note that in the large $\Gamma$ limit Eq.~(\ref{eq:jointProbability}) takes a large deviation form \cite{touchette}. Since $\Gamma$ multiplies time in $\xi^+_{\text{NI}}$, we rescale time as $\tilde{\tau}= \tau \Gamma$. The corresponding rescaled stochastic equation for $\xi^+_i$ is
\begin{align}\label{eq:smallNoiseXp}
\dot{\xi}^+_i(\tilde{\tau}) =  \frac{1}{2}(1-{\xi^+_i}^2) +\epsilon\xi^+_i \sum_j O_{ij} \phi_j , 
\end{align}
where we have defined the noise strength $\epsilon \equiv 1/\Gamma$. The limit $\Gamma\rightarrow \infty$ is therefore equivalent to the small-noise limit of~(\ref{eq:smallNoiseXp}).  
Stochastic differential equations in the limit of small noise are described by the Freidlin-Wentzell (FW) large deviation theory~\cite{freidlinWentzell,demboZeituni}: $\xi^+_i$ obeys a large deviation principle (LDP) , with rate $\epsilon^{-2}$ and rate function $\mathcal{I}[\xi^+] \equiv \epsilon^2 I[\xi^+]$. In this small-$\epsilon$ limit, the trajectories $\xi^+_i$ are approximately Gaussian distributed around the likeliest trajectory $\xi^+_{\text{NI}}$ \cite{touchette}:
\begin{equation}\label{eq:largeDeviationGaussian}
P[\xi^+_i] \sim e^{-\frac{\epsilon^{-2}}{2}\int_0^{\tilde{\tau}} \sum_{ij}\mathcal{I}^{(2)}_{ij} [\xi^+_i(\tilde{\tau}^\prime) - \xi^+_{NI}(\tilde{\tau}^\prime)] [\xi^+_j(\tilde{\tau}^\prime) - \xi^+_{NI}(\tilde{\tau}^\prime)] \mathrm{d}\tilde{\tau}^\prime } ,
\end{equation} 
where the second variation $\mathcal{I}^{(2)}_{ij}$ is given by
\begin{equation}
\mathcal{I}^{(2)}_{ij}\equiv \frac{\delta^2 \mathcal{I}}{\delta \xi^+_i(\tilde{\tau}^\prime)\delta \xi^+_j(\tilde{\tau}^\prime)}\Big\vert_{\xi^+_{NI}} .
\end{equation}
Thus, trajectories that deviate significantly from the non-interacting limit are exponentially suppressed.
The large deviation formalism also applies to real time evolution, where again the dominant trajectory is given by the deterministic result $\xi^+_{\text{NI}}(t)$. However, in contrast to $\xi^+_{\text{NI}}(\tau)$, $\xi^+_{\text{NI}}(t)$ has an infinite number of singularities as a function of time \cite{nonEquilibrium}. This leads to a breakdown of the expansion about the non-interacting saddle point, which can be expected to have consequences for sampling. Even for large $\Gamma$, regions in time that are close to the singularities in the saddle point trajectory are expected to be associated with enhanced fluctuations, leading to difficulties in sampling. This observation may lie at the root of the enhanced fluctuations of the disentangling variables found in the vicinity of dynamical quantum phase transitions \cite{heyl2013,heyl2018}, reported in \cite{stochasticApproach,nonEquilibrium}. 

\section{Saddle Point Equation}\label{app:saddlePoint}
In this Appendix, we provide details on the saddle point equation discussed in Section~\ref{sec:findingSP}, including its numerical solution, its generalization to other observables and the higher dimensional case. We also discuss a toy model of an integral for which several saddle points exist, and provide further details on our discussion of quantum phase transitions considering the quantum Ising chain as a concrete example. 

\subsection{Numerical Solution}\label{app:recursiveSolution}
The saddle point equation~(\ref{eq:loschTISP}) for the Loschmidt amplitude can be solved recursively, exploiting the intuition that the saddle point field configuration $\varphi_{\text{SP}}(\tau^\prime) \equiv \varphi_{\text{SP}}(\tau^\prime|\tau_f)$ should change little if $\tau_f$ is increased by a small amount $\Delta t$.
In practice, one assumes
\begin{equation}
\varphi_{\text{SP}}(\tau^\prime | \tau_f + \Delta t ) \approx \varphi_{\text{SP}}(\tau^\prime | \tau_f) 
\end{equation}
for $\tau^\prime < \tau_f + \Delta t$. The field $\varphi_{\text{SP}}(\tau^\prime | \tau_f) $ is then used to compute $\xi^{+}_i\vert_{\text{SP}}$ and $\Xi_{ij}\vert_{\text{SP}}$. Using these quantities, one can in turn produce a better approximation of $\varphi_{\text{SP}}(\tau^\prime|\tau_f + \Delta t )$ according to the saddle point equation~(\ref{eq:loschTISP}). This procedure can be iterated until the field configuration has converged to a desired level of accuracy. The convergence of the recursion is determined by defining a quantity $\varepsilon$ which measures how much the approximate saddle point field varies after an iteration of the algorithm. A suitable definition is
\begin{equation}\label{eq:convParam}
\varepsilon \equiv \frac{1}{k} \sum_{m=1}^k \big| \bar{\varphi}_{\text{SP}}(\tau_m| \tau_f+\Delta\tau) -  \varphi_{\text{SP}}(\tau_m| \tau_f+\Delta\tau) \big|
\end{equation}
where $\varphi_{\text{SP}}$  and $\bar{\varphi}_{\text{SP}}$ are the old and updated estimates of the SP field respectively, evaluated at the discrete times $\tau_m$. Convergence is then defined as $\varepsilon< \varepsilon^*$, where $\varepsilon^*$ is a threshold of choice. 
The runtime of this recursive algorithm scales quadratically with the number of time steps $n$; this is because for each $1<k<n$ one needs to perform $k$ calculations in order to compute $\xi^{+}_i\vert_{\text{SP}}$, so that summing over all $k$ the total number of calculations to perform is of order $n(n+1)/2$. In principle, the computational cost is further increased by having to repeat each step multiple times to attain convergence. However, for reasonable values of the threshold $\varepsilon^*$, numerical evaluation shows that the recursive algorithm has rapid convergence, typically requiring only $1-2$ iterations.
\begin{figure}
\centering
  \includegraphics[width=.9\linewidth]{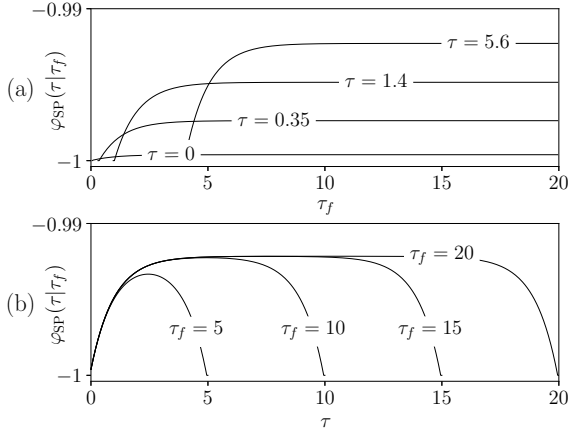}
  \caption{Behavior of the saddle point field $\varphi_{\text{SP}}(\tau|\tau_f)$ obtained from the recursive solution of the saddle point equation (\ref{eq:loschTISP}) for a quantum Ising chain with $\Gamma=\Gamma_c/2$, initialized in the $\ket{\Downarrow}$ state. (a) At times $\tau\ll \tau_f$, for sufficiently large stopping time $\tau_f$, the saddle point field $\varphi_{\text{SP}}(\tau|\tau_f)$ attains a $\tau_f$-independent value and can be considered to have converged. (b) At short times $0 \lesssim \tau$, we observe a transient behavior in the SP field, which depends on initial conditions and corresponds to the imaginary time evolution of the initial state towards the ground state. At late times $\tau \lesssim \tau_f$, the SP field is affected by the constraint $\varphi_{\text{SP}}(\tau_f|\tau_f)=-1$.  For intermediate times $0 \ll \tau \ll \tau_f$ the saddle point field attains a plateau value $\varphi_P$, which gives the main contribution to observables as $\tau_f\rightarrow \infty$.}
\label{fig:phiSSlosch}
\end{figure}
From recursively solving the SP equation, we find that for sufficiently large $\tau_f$ the value $\varphi_{\text{SP}}(\tau^\prime| \tau_f)$ with $\tau^\prime \ll \tau_f$ no longer changes with $\tau_f$, settling to a value $\varphi_{\text{SP}}(\tau^\prime|\infty) \equiv \varphi_{\text{SP}}(\tau^\prime)$; this is illustrated in Fig.~\ref{fig:phiSSlosch}(a). 
Because of this, when recursively solving the SP equation one only needs to update the SP configuration at the times $\tau^\prime$ such that $\varphi_{\text{SP}}(\tau^\prime|\tau_f) \neq \varphi_{\text{SP}}(\tau^\prime)$ to a desired level of precision; this speeds up the recursive solution significantly. The SP equation~(\ref{eq:loschTISP}) prescribes that the value of the saddle point field at the end time is always $\varphi_{\text{SP}}(\tau_f| \tau_f)=-1$. Thus, the saddle point field $\varphi_{\text{SP}}(\tau^\prime | \tau_f)$ cannot attain a steady state, i.e. for finite $\tau_f$ there exists no time scale $\tau_{SS}$ such that $\partial_{\tau^\prime}\varphi_{\text{SP}}(\tau^\prime, \tau_f) \approx 0$ $\forall$ $\tau^\prime> \tau_{SS}$. However, the numerical results show that for sufficiently large $\tau_f$ the SP field $\varphi_{\text{SP}}(\tau^\prime,\tau_f)$ attains a plateau value at times $0 \ll \tau^\prime \ll \tau_f$; this is illustrated in Fig.~\ref{fig:phiSSlosch}(b). The extent of the plateau grows as $\tau_f$ is increased; since the action is extensive in time, the plateau value provides the dominant contribution to observables in the large $\tau_f$ limit. The plateau value of $\varphi_{\text{SP}}$ can be found analytically, as discussed in Section~\ref{sec:findingSP}; the analytical results are in perfect agreement with the numerical solution.

\subsection{Saddle Point for General Observables}
In the disentanglement formalism, the Euclidean time evolution of an observable $\mathcal{O}$ is given by 
\begin{align}\label{eq:generalObsApp}
\bra{\psi_0} \hat{U}(\tau) \hat{\mathcal{O}} \hat{U}(\tau) \ket{\psi_0} = \int \mathcal{D} \phi\, e^{-S_0[\phi]} F_{\mathcal{O}}[\phi] ,
\end{align}
where $\phi= \{ \phi_{f,i}^a, \phi_{b,i}^a \}$ collectively denotes the two sets of HS fields introduced to decouple the two time-evolution operators, and the classical function $F_{\mathcal{O}}$ is given by Eq.~(\ref{eq:classicalFunction}). As discussed in Section~\ref{sec:effectiveAction}, the trajectory yielding the largest contribution to the integral can be found by extremizing the effective action
\begin{align}\label{eq:effectiveActionApp}
S_\mathcal{O} \equiv  S_0[\phi] - \log f_\mathcal{O}[\phi] .
\end{align}
Consider the normalization function, corresponding to setting $\hat{\mathcal{O}}=\mathbbm{1}$ in~(\ref{eq:generalObsApp}) and given by Eq.~(\ref{eq:normalization}). For the 1D quantum Ising model, the effective action for this quantity is given by
\begin{widetext}
\begin{align}\label{eq:normalizationAction}
S_{\mathbbm{1}} =\frac{1}{2} \int_0^{\tau_f} \mathrm{d}\tau \sum_i \left[ \frac{J}{2} \sum_j \mathcal{J}^{-1}_{ij} [\varphi_{f,i}  \varphi_{f,j}+\varphi^*_{b,i} \varphi^*_{b,j}]  - \Gamma \xi^+_{f,i} - \Gamma \xi^{+*}_{b,i}+ J \varphi_{f,i}+ J \varphi^*_{b,i} \right]  - \sum_i \log\left[ 1+ \xi^+_{f,i}(\tau_f) \xi^{+*}_{b,i}(\tau_f) \right].
\end{align}
\end{widetext}
By varying Eq.~(\ref{eq:normalizationAction}), we obtain the SP equations for the normalization: 
\begin{align}\label{eq:SPeqnNormalization}
\begin{split}
& J \sum_j \mathcal{J}^{-1}_{ij} \varphi_{f,j}(\tau^\prime) \vert_{\text{SP}} = \\
&\,\, \Gamma\int_0^{\tau_f} \Xi_{f,i}(\tau,\tau^\prime) \dd\tau  \vert_{\text{SP}} - J + \frac{2 \xi^{+*}_{b,i} \Xi_i(\tau_f,\tau^\prime)  }{1+ \xi^+_{f,i}(\tau_f) \xi^{+*}_{b,i}(\tau_f)} \Big\vert_{\text{SP}} .
\end{split}
\end{align}
The same equation is satisfied by $\varphi_{b,i}\vert_{\text{SP}}$, with the replacement $f\leftrightarrow b$. 
By direct substitution, one readily verifies that the plateau of the Loschmidt amplitude SP is a fixed point of Eq.~(\ref{eq:SPeqnNormalization}) at all times. Thus, choosing the mean field ground state as the initial state eliminates both the transient and the late-time behavior of Eq.~(\ref{eq:SPeqnNormalization}) (in contrast, for any initial state, the solution of Eq.~(\ref{eq:loschTISP}) deviates from the plateau at late times due to the boundary condition $\phi(\tau_f)=-1$). More generally, local observables expressed in a translationally invariant way correspond to stochastic functions 
\begin{align}\label{eq:functionGeneralObs}
f_\mathcal{O}(\tau_f) = f_\mathcal{\mathbbm{1}}(\tau_f)  \sum_i\bar{f}_{\mathcal{O},i}(\tau_f)  ,
\end{align} 
where $\bar{f}_{\mathcal{O},i}(\tau_f)$ is a function of $\xi^+_{f,i}(\tau_f)$ only. For instance, for the magnetization one has $\bar{f}_{\mathcal{M},i}= (1-\xi^+_{f,i}\xi^{+*}_{b,i})/(1+\xi^+_{f,i}\xi^{+*}_{b,i})$; see Eq.~(\ref{eq:longitudinalMagnetization}). It can be readily seen that the SP equation obtained by extremizing the effective action for (\ref{eq:functionGeneralObs}) differs by Eq.~(\ref{eq:SPeqnNormalization}) by a term proportional to $1/N$. Furthermore, the extra term is also proportional to $\Xi_{\text{SP}}(\tau_f,\tau)$; at the plateau, one has $\Xi_P(\tau_f,\tau) \propto e^{-(\Gamma \xi^+_P - J \phi_P)(\tau_f-\tau)}$,  so that the extra term is inconsequential as $\tau_f\rightarrow \infty$. 
Thus, the SP equation for any local observable differs from~(\ref{eq:SPeqnNormalization}) by a term which is suppressed both as $\tau_f \rightarrow \infty$ and as $N\rightarrow \infty$. This implies that the plateau SP trajectory for the Loschmidt amplitude can be used for all other ground state expectation values, both for analytical and numerical applications.

\subsection{Higher Dimensions}\label{app:higherDimensions}
For the $D$-dimensional quantum Ising model, the Euclidean SDEs are given by
\begin{subequations}\label{eq:highDising}
\begin{align}
\dot{\xi}^+_{\boldsymbol{i}} &= \frac{\Gamma}{2} (1-{\xi^+_{\boldsymbol{i}}}^2) + \xi^+_{\boldsymbol{i}} \sum_{d=1}^D J_d \varphi^d_{\boldsymbol{i}}  ,\\
\dot{\xi}^z_{\boldsymbol{i}} &= -\Gamma \xi^+_{\boldsymbol{i}} +  \sum_{d=1}^D J_d \varphi^d_{\boldsymbol{i}} , \\
\dot{\xi}^-_{\boldsymbol{i}} &= \frac{\Gamma}{2} \exp {\xi}^z_{\boldsymbol{i}} ,
\end{align}
\end{subequations}
with multicomponent indices $\boldsymbol{i}=\{ i_{1}, \dots, i_{D} \} $.
The Euclidean Loschmidt amplitude is given by
\begin{align}
A(\tau_f) =\int \mathcal{D}\varphi e^{-S_0[\varphi] - \frac{1}{2}\int_0^{\tau_f} \sum_{\boldsymbol{i}} \mathrm{d}\tau \left[ \sum_{d=1}^D J_d \varphi^d_{\boldsymbol{i}}(\tau) - \Gamma \xi^+_{\boldsymbol{i}}(\tau)  \right] }.
\end{align}
The saddle point equation obtained by varying the effective action with respect to $\varphi^d_{\boldsymbol{i}}(\tau^\prime)$ is then given by
\begin{align}\label{eq:SpeqD}
\varphi^d_{\boldsymbol{i}}(\tau^\prime) \vert_{\text{SP}} = \frac{\Gamma}{J_d} \sum_{j_d}  \mathcal{J}^d_{i_d j_d} \int_0^{\tau_f} \Xi^d_{i_1 \dots j_d \dots i_D} (\tau,\tau^\prime) \big\vert_{\text{SP}} \mathrm{d}\tau  - 1 .
\end{align}
The functional derivative $\Xi^d{\boldsymbol{i}}(\tau,\tau^\prime)$ can be obtained by varying the equation of motion of $\xi^+_{\boldsymbol{i}}$, as in the one-dimensional case:
\begin{align}
\Xi^d_{\boldsymbol{i}}(\tau, \tau^\prime) =J_d \xi^+_{\boldsymbol{i}}(\tau^\prime) \theta(\tau -\tau^\prime) e^{ \int_{\tau^\prime}^\tau  \mathrm{d}s \left[  - \Gamma \xi^+_{\boldsymbol{i}} (s) + \sum_{d=1}^D J_d \varphi^d_{\boldsymbol{i}}(s)  \right] }.
\end{align}
For a translationally invariant system one has $\xi_{\boldsymbol{i}}\vert_{\text{SP}} = \xi^+_{\text{SP}} $, $\Xi^d_{\boldsymbol{i},\boldsymbol{j}}\vert_{\text{SP}} = \Xi^d_{\text{SP}}$, $\varphi^d_{\boldsymbol{i}}\vert_{\text{SP}}=\varphi^d_{\text{SP}}$, such that the SP equation simplifies to
\begin{align}\label{eq:TISP_D}
\varphi^d_{\text{SP}}(\tau^\prime) = \frac{ \Gamma }{J_d} \int_{0}^{\tau_f}  \Xi^d_{\text{SP}} (\tau,\tau^\prime) \big\vert_{\text{SP}} \mathrm{d}\tau  - 1 .
\end{align}
For a fully isotropic system with $J_1 = \dots = J_D=J$, one additionally has $
\varphi^d_{\text{SP}}=\varphi_{\text{SP}}$ and the SP equations further simplify to 
\begin{align}
\varphi_{\text{SP}}(\tau^\prime) &= \frac{ \Gamma}{J} \int_{0}^{\tau_f}  \Xi_{\text{SP}} (\tau,\tau^\prime) \big\vert_{\text{SP}} \mathrm{d}\tau  - 1 ,\label{eq:TRISP_D} \\
\Xi_{\text{SP}}(\tau,\tau^\prime) & = \theta(\tau-\tau^\prime) \xi^+_{\text{SP}}(\tau^\prime) e^{\int_{\tau^\prime}^\tau \mathrm{d}s \left[ -\Gamma \xi^+_{\text{SP}}(s) + D J \varphi_{\text{SP}}(s)  \right] } ,
 \\
\dot{\xi}^+_{\text{SP}} &=  \frac{\Gamma}{2} \left( 1- \xi^{+2}_{\text{SP}} \right) + D J \xi_{\text{SP}}^+ \varphi_{\text{SP}} .\label{eq:TRISP_D_xp}
\end{align}
Plateau equations can be derived from Eqs~(\ref{eq:TRISP_D}) and (\ref{eq:TRISP_D_xp}) in the large $\tau_f$ limit:
\begin{subequations}\label{eq:SPscaleinv}
\begin{align}
\varphi_P  &=  \frac{DJ \varphi_P}{\Gamma \xi^+_P - D J \varphi_P} ,  \\
\varphi_P &= - \frac{\Gamma}{2DJ}\frac{1-\xi^{+2}_{P}}{\xi^+_P} .
\end{align} 
\end{subequations}
These equations are solved by~(\ref{eq:PsolutionsD}).

\subsection{Multiple Saddle Points: Toy Example}\label{app:SPintegrals}
\begin{figure}
\centering
  \includegraphics[width=.9\linewidth]{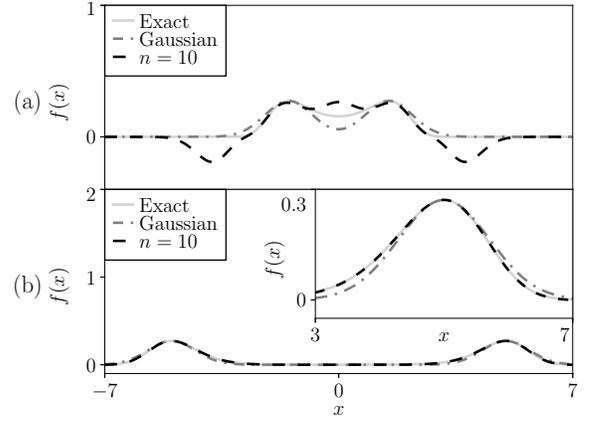}
  \caption{Expansions of integrals in the presence of more than one saddle point. We show the integrand $f(x)$ defined in Eq.~(\ref{eq:toyIntegrand}), comparing the exact value (full line), the approximation corresponding to truncating Eq.~(\ref{eq:toyExpansion}) to Gaussian order (dash-dotted line), and the approximation obtained from Eq.~(\ref{eq:toyExpansion}) with $n=10$ (dashed line). (a) For $a=1.5$, the saddle points at $x_{\text{SP}}=\pm a$ are close to each other and the expansion~(\ref{eq:toyExpansion}) produces a worse approximation to the integral for $n=10$ ($37 \%$ error) than for the Gaussian approximation $n=2$ ($3 \%$ error). (b) For $a=5$, the saddle points are well separated and the higher order expansion closely approximates the integrand, as shown in the inset. This leads to a better performance for the $n=10$ approximation, which gives the correct integral within  $1.4 \%$, compared to an error of $3.7 \%$ for the Gaussian approximation. For the present example, both expansions eventually break down as $n$ is increased due to their asymptotic nature. }\label{fig:spToyExample}
\end{figure}
Here we provide a toy example illustrating the expansion of an integral which has two different saddle points. We consider the integral
\begin{align}
I(a) &= \int_{-\infty}^\infty f(x)  \dd x , \label{eq:toyIntegral} \\
f(x) &= \mathcal{C}_a e^{-S(x)}, \label{eq:toyIntegrand}\\
S(x) &= \frac{x^4}{4a^2} - \frac{x^2}{2}, \label{eq:toyAction} 
\end{align}
where $\mathcal{C}_a$ is a normalization constant defined by $I(a)=1$. Extremization of $S(x)$ with respect to $x$ yields two minima, $x_{\text{SP}} = \pm a $. One can then expand the action around each SP as
\begin{align}
S = S_{\text{SP}} + \frac{1}{2!} S^{(2)}_{\text{SP}} (x-x_{\text{SP}})^2 + S^h ,
\end{align}
where $S^{(2)}$ is the second variation evaluated at the SP and $S^h$ includes all contributions of higher order. We then
approximate Eq.~(\ref{eq:toyIntegral}) as
\begin{align}\label{eq:toyExpansion}
\begin{split}
& I(a) \approx \\
&\mathcal{C}_a \sum_{\text{s.p.}} e^{-S_{\text{SP}}}\hspace{-1.5mm}  \int_{-\infty}^{\infty} \hspace{-1.5mm} e^{-\frac{1}{2} S^{(2)}_{\text{SP}}(x-x_{\text{SP}})^2} [ 1+\sum_{m=3}^{n} \alpha_m (x-x_{\text{SP}})^m ] \dd x ,
\end{split}
\end{align}
where the coefficients $\alpha_m$ are obtained by Taylor expanding $e^{S^h}$ and the leftmost sum runs over the two saddle points. For $n \leq 2$, none of the $\alpha_m$ is included and thus Eq.~(\ref{eq:toyExpansion}) reduces to the evaluation of Gaussian fluctuations around the SP. To show how well the approximation~(\ref{eq:toyExpansion}) captures the true value of $I(a)$, in Fig.~\ref{fig:spToyExample} we compare the exact and approximate integrands for different values of $a$, $n$. 
We find that the approximation~(\ref{eq:toyExpansion}) for fixed $n$ gets more accurate as $a$ increases, such that the SPs are better spaced out. This is an example of the ``small overlap" condition discussed in the main text: one can separately expand about the two saddle points and add up the individual contributions of the expansions, provided that regions (in this case, along the $x$ axis) which contribute significantly to one integral give negligible contribution to the other.

\subsection{Quantum Phase Transitions}
\begin{figure}[t]
\centering
\includegraphics[width=\linewidth, trim=0 0 0 30, clip]{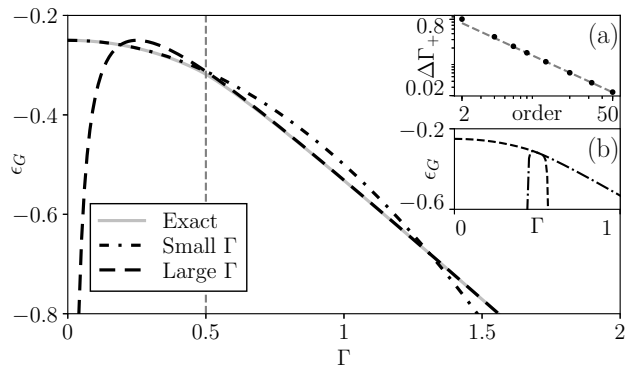}\caption{Crossing of different perturbative series for the quantum Ising chain in the thermodynamic limit $N\rightarrow \infty$. The main panel shows the exact ground state energy density (full gray line) as a function of the transverse field $\Gamma$ for $J=1$; this is compared to the approximate values obtained by perturbatively expanding the exact result to second order in small $\Gamma$ (dashed line) or large $\Gamma$ (dash-dotted line). The small-$\Gamma$ and large-$\Gamma$ expansions cross at three points $\Gamma_1<\Gamma_2<\Gamma_3$, where $\Gamma_2=0.5 = \Gamma_c$ is the critical point of the model. As the order of the perturbative expansions is increased, $\Delta \Gamma_+ \equiv \Gamma_3-\Gamma_2$ approaches zero as a power law, as shown in panel (a): the dashed gray line shows the power law fit. The same applies to $\Delta \Gamma_- \equiv \Gamma_2-\Gamma_1$ (not shown). Furthermore, the small-$\Gamma$ and large-$\Gamma$ series rapidly diverge for $\Gamma \gtrsim 0.5$ and $\Gamma \lesssim 0.5$ respectively: this is illustrated in panel (b), where we show the 100-th order expansions of $\epsilon_G$ in small and large $\Gamma$. These observations corroborate the picture proposed in the main text: within the present field theoretical description, quantum phase transitions in spin chains can be understood as arising from an abrupt switch in which saddle point expansion dominates in the thermodynamic limit. In the present case, for each value of $\Gamma$ one series is discarded since it is divergent, while the other provides the correct result.}
\label{fig:crossingSeries}
\end{figure}
In Section~\ref{sec:fullExpansion}, we provided a general discussion of how quantum phase transitions emerge from the field theoretical application of the disentanglement approach. Namely, due to Eq.~(\ref{eq:gsEnergyAssumption}), a quantum critical point corresponds to the value for which there is an abrupt change in the dominant contribution to the grand state energy. This can be visualized for the quantum Ising chain, for which an exact analytical solution is available~\cite{Pfeuty1970}.
In Fig.~\ref{fig:crossingSeries}, we show the small-$\Gamma$ and large-$\Gamma$ perturbative expansions of the ground state energy (\ref{eq:GSenergyPfeuty}), given by (\ref{eq:exactPerturbativeSeries}). For any finite order in perturbation theory, the series cross at three points. As the order of both expansions is increased, the three crossing points converge towards a single point, the critical point $\Gamma_c=J/2$; see panel (a). Panel (b) further highlights that the small-$\Gamma$ expansion is divergent for $\Gamma>\Gamma_c$; therefore, the relative terms will not contribute in this regime. Similarly, the large-$\Gamma$ series does not contribute when $\Gamma<\Gamma_c$.

\section{Mean Field Approximation}\label{app:meanField}
In Section~\ref{sec:leadingOrder} we discussed the relation between mean field (MF) theory and the disentanglement method, showing that MF corresponds to the leading order of a more general expansion.
To aid comparison with the results of the main text, here we outline the derivation of the MF ground state for the $D$-dimensional quantum Ising model~(\ref{eq:Ising}). The MF approach consists in approximating the ground state by the product state which minimizes the energy of the system. The ground state is thus parameterized via the variational ansatz
\begin{align}
\ket{\text{MF}}= \otimes_i ( \cos\theta \ket{\uparrow}_i + \sin\theta \ket{\downarrow}_i).
\end{align}
This ansatz gives a ground state energy density
\begin{align}
\epsilon_\text{MF}(\theta) =  - \frac{\Gamma}{2} \sqrt{1-\cos(2\theta)^2} - \frac{1}{4}  J D \cos(2\theta)^2 .
\end{align}
Minimizing this with respect to $x\equiv \cos(2\theta)$, one gets three solutions:
\begin{subequations}\label{eq:MFsolutions}
\begin{align}
x&=\pm \frac{\sqrt{D^2 J^2 - \Gamma^2}}{ DJ }, \\
x&=0 ,
\end{align}
\end{subequations}
where the first solution is only valid for $\Gamma<DJ$. For each value of $\Gamma$, one then chooses the solution in~(\ref{eq:MFsolutions}) which minimizes $\epsilon$. This yields the mean-field approximation to the ground state energy density:
\begin{align}
\epsilon_{\text{MF}}=
\begin{cases} 
 -\frac{D^2 J^2 + \Gamma^2}{4 D J}  \quad &\text{for}  \, \Gamma<DJ ,\\
 -\frac{\Gamma}{2} \quad &\text{for}  \, \Gamma\geq DJ .
\end{cases}
\end{align}
Within the MF approximation, the ground state magnetization is then given by
\begin{align}
m_{\text{MF}}=
\begin{cases}
 \pm \frac{\sqrt{(D J-\Gamma )(DJ+\Gamma)}}{2 DJ}  \quad &\text{for}  \, \Gamma<DJ, \\
 0 \, &\text{for}  \, \Gamma \geq DJ .
\end{cases}
\end{align}
The MF approximation predicts a quantum phase transition at $\Gamma_c^{MF}= DJ$. The same results may be obtained by writing
$\hat{S}^z_i = m^z + \delta\hat{S}^z_i$ and neglecting quadratic fluctuations, $ \delta\hat{S}_i^z \delta\hat{S}_j^z \approx 0$. The definition $m_z \equiv \langle \hat{S}^z_i \rangle $ then gives a self-consistency condition.
The MF results provided in this Section correspond to the SP result given in the main text; in particular, $\varphi_P = 2D m_{\text{MF}}$ is precisely the effective field felt by each spin (i.e. the \textit{mean field}).
\end{document}